\documentclass{article}

\usepackage{arxiv}

\usepackage[utf8]{inputenc} 
\usepackage[T1]{fontenc}    
\usepackage{hyperref}       
\usepackage{url}            
\usepackage{booktabs}       
\usepackage{amsfonts}       
\usepackage{nicefrac}       
\usepackage{microtype}      
\usepackage{lipsum}		
\usepackage{graphicx}
\usepackage{natbib}
\usepackage{doi}
\usepackage{graphicx}
\usepackage{newtxtext}
\usepackage{newtxmath}
\usepackage{natbib}
\usepackage{multirow}
\usepackage{booktabs}
\usepackage{lipsum}

\usepackage{hyperref}
\hypersetup{
    colorlinks = true,
    urlcolor   = blue,
    citecolor  = black,
}

\newcommand{\RomanNumeralCaps}[1]

\title{Effects of Plunging Acceleration on the Passive Morphing of Avian-Inspired Flexible Foils}

\date{} 					

\author{ \href{}{Hibah Saddal} \\
    Aerospace Engineering \\
    School of Metallurgy and Materials \\
    University of Birmingham \\
    Birmingham, UK \\
    \texttt{hxs619@student.bham.ac.uk} \\
	\And
	\href{}{Lucky Babu Jayswal} \\
	Aerospace Engineering \\
    School of Metallurgy and Materials \\
    University of Birmingham \\
    Birmingham, UK \\
	\texttt{luckyjayswal54@gmail.com} \\
	\AND
	\href{https://orcid.org/0000-0001-5166-9968}{Chandan Bose}\thanks{Corresponding author}\\
	Aerospace Engineering \\
    School of Metallurgy and Materials \\
    University of Birmingham \\
    Birmingham, UK \\
	\texttt{c.bose@bham.ac.uk} \\
}

\renewcommand{\shorttitle}{Effects of Plunging Acceleration on the Passive Morphing of Avian-Inspired Flexible Foils}

\hypersetup{
pdftitle={Effects of Plunging Acceleration on the Passive Morphing of Avian-Inspired Flexible Foils},
pdfauthor={Hibah Saddal, Lucky Babu Jayswal, Chandan Bose},
pdfkeywords={Fluid-structure interaction, passive morphing, accelerated plunging, flexible foils, transient aerodynamics, bio-inspired propulsion.},
}

\title{Effects of Plunging Acceleration on the Passive Morphing of Avian-Inspired Flexible Foils}

\begin{document}
\maketitle

\begin{abstract}
This study investigates the dynamics of passively morphing foils under accelerated plunging, establishing mechanistic links between transient kinematics, structural compliance, and aerodynamic performance. Two-way coupled simulations are performed for three wing geometries: a symmetric NACA0012 foil and two bio-inspired geometries based on falcon and owl wing sections, across non-dimensional bending rigidity values ($K_B = 5$ -- $100$), chordwise flexible segment extents from the trailing-edge ($25\%$, $50\%$, and $75\%$), and transition speed parameters ($a_s = 3$ -- $11$). The present findings reveal that flexible trailing-edge configurations exhibit improved aerodynamic performance relative to stiffer foils, and the aerodynamic benefit of trailing-edge compliance is strongly influenced by wing geometry. A geometry-specific optimal bending stiffness exists: $K_B = 10$ for the NACA0012 foil and owl wing section, and $K_B = 7.5$ for the falcon wing section, beyond which additional flexibility degrades performance. The extent of the chordwise flexible segment critically governs the aeroelastic response. Whilst a $25\%$ flexible segment produces behaviour indistinguishable from a rigid wing, extending flexibility to $75\%$ of the chord induces highly unsteady lift fluctuations, particularly for the NACA0012 foil, for which the $C_{l_{RMS}}$ increases sharply. The bio-inspired foils, in contrast, exhibit a moderate reduction in $C_{l_{RMS}}$ for the 50\% and 75\% cases, reflecting the stabilising influence of their cambered geometry. Increasing $a_s$ from $3$ to $11$ monotonically amplifies trailing-edge deflection, strengthens the leading- and trailing-edge vortices, and intensifies the coupling between structural deformation and instantaneous lift. These findings provide new physical insight into bio-inspired propulsion and manoeuvring strategies, with implications for the design of passively adaptive lifting surfaces in unsteady environments.
\end{abstract}


\section{Introduction}
\label{sec:intro}

Flexible lifting surfaces have attracted considerable research attention owing to their ability to adapt passively under aerodynamic loading, offering potential advantages in efficiency, disturbance rejection, and load alleviation compared to their rigid counterparts \citep{kang2011effects, martinez2024unsteady, otomo2025scaling, cao2025gust, micklem2026harnessing}. In biological flyers, this compliance is not incidental but functionally integrated: avian wings exhibit spatially distributed stiffness gradients that enable continuous shape adaptation even without active actuation \citep{klaassen2020passive}. This allows birds to sustain stable flight, recover from gust-induced perturbations, and execute agile manoeuvres through coordinated morphing of wing and tail configurations \citep{harvey2022birds, harvey2022gull, martinez2024steady}. The aerodynamic consequences of this passive compliance motivate the present study, which seeks to characterise the fluid-structure interaction (FSI) dynamics of bio-inspired morphing foils under transient kinematics. The findings of this study will be particularly relevant to futuristic unmanned aerial vehicles (UAVs) and drones, operating in gusty environments \citep{harvey2022review, jeger2024adaptive, wuest2024agile}.

The effects of trailing-edge flexibility on aerodynamic efficiency have been demonstrated by several recent studies. \citet{boughou2024investigation} reports enhanced aerodynamic efficiency for stationary cambered foils with a flexible trailing-edge at low angles of attack and moderate-to-high Reynolds numbers, attributing the improvement to deformation-mediated modification of the pressure distribution. Numerical investigations of segmented flexible NACA0012 aerofoils by \citet{hefeng2015numerical} also demonstrate substantial lift augmentation relative to rigid baselines. However, the findings of these studies are limited to moderately stiff foils under stationary conditions, exhibiting very low deformation. Furthermore, the analysis is restricted to static angle-of-attack sweeps without unsteady kinematic forcing, implying that the reported lift augmentation mechanism is limited only to attached or mildly separated steady flow and cannot be extended to the transient vortex-dominated regimes characteristic of accelerating motions. 

Recently, \citet{liu2024unsteady} has studied a single-degree-of-freedom elastically supported rigid NACA0012 aerofoil subjected to a ramp gust. The authors demonstrate that passive pitch can suppress up to $\sim 80\%$ of gust-induced lift fluctuations by appropriately placing the pitching axis upstream. \citet{cao2025gust} shows that passive deformation of flexible trailing-edge NACA0012 airfoils attenuates periodic gust-induced loading through coupled deformation-vortex interaction.  \citet{otomo2025scaling} experimentally demonstrates that unsteady lift alleviation in plunging aerofoils with flexible trailing-edges scales with a product of two Cauchy numbers, one based on the freestream velocity and one on the plunging velocity. This study provides a non-dimensional framework that collapses both trailing-edge deflection and load alleviation data across a wide range of kinematic and flexibility conditions. Extending beyond purely passive compliance, \citet{micklem2026harnessing} demonstrates that integrating a capacitive e-skin with active camber morphing in a soft robotic wing enables a hybrid passive-active disturbance rejection strategy, significantly reducing the unwanted lift impulse relative to a comparable rigid wing. Note that all these studies are conducted on geometrically simple, symmetric, or mildly cambered profiles with uniform or segmented chordwise stiffness distributions, and none incorporates the real avian wing geometries. Furthermore, none of these works considers transient or accelerated kinematics; the reported force and deformation responses are therefore valid only for harmonic or stationary conditions and cannot be extended to the non-periodic forcing that characterises realistic manoeuvres.  

In the harmonically oscillating kinematics, characterised by a fixed frequency and amplitude, the aerodynamic response is well described by classical non-dimensional parameters, such as the dynamic plunge velocity, reduced frequency, and Strouhal number \citep{gursul2019plunging, bose2021dynamic, majumdar2022transition}. In contrast, realistic flight manoeuvres and gust-driven responses involve strongly transient kinematics in which instantaneous acceleration, rather than frequency, governs the rate of vorticity production at the surface, the evolution of the feeding shear layer, and the timescale over which a leading-edge vortex (LEV) attains saturation and detaches \citep{bull2021unsteady}. Under accelerated motion, the vortex dynamics, including shear-layer instability, LEV growth and convection, as well as the resulting wake topology, are fundamentally altered \citep{sattari2012growth}. Acceleration, therefore, is an important governing parameter for understanding transient FSI effects, yet it has received little systematic attention in the context of passively deforming wings.

This omission raises a physically non-trivial question. Under harmonic kinematics, structural compliance at intermediate stiffness is generally associated with enhanced force generation through passive pitch adaptation and camber modulation \citep{heathcote2007flexible, shyy2010recent, kang2011effects}, though this benefit diminishes or reverses at excessive flexibility \citep{shah2024controlling, eldredge2010roles}. Under transient acceleration, however, the structural response timescale may be commensurate with or shorter than the vortex formation timescale, introducing a qualitatively different coupling between wing deformation and vortical flow evolution. Rapid, large-amplitude deformations arising from impulsive loading could geometrically perturb the feeding shear layer, potentially modifying the rate of circulation growth or the timescale to LEV saturation in ways that differ qualitatively from the harmonic case \citep{rival2014characteristic, mulleners2017flow}. Whether the favourable deformation-mediated force augmentation observed under periodic kinematics persists under transient acceleration, or whether the altered vortex dynamics produce a decoupling between deformation amplitude and force generation, is therefore not evident a priori and has not been previously examined.

Furthermore, despite biological evidence that avian wings exhibit pronounced chordwise stiffness gradients due to varying thickness, the majority of existing studies on flexible lifting surfaces considered flat plates or conventional symmetric aerofoils with a single, globally prescribed stiffness \citep{heathcote2007flexible, eldredge2010roles, kang2011effects}. This simplification, though analytically tractable and computationally convenient, represents a significant departure from the structural reality of biological wings, in which a stiff leading-edge structure gradually transitions to a highly compliant trailing-edge region. Biological measurements have established that chordwise flexural stiffness in flight appendages varies by one to two orders of magnitude across the chord  \citep{combes2003flexural2, combes2003flexural1, klaassen2020passive}. The location and spatial extent of compliant regions along the chord govern both the camber profile that develops under aerodynamic loading and the chordwise position of deformation-induced passive pitch adaptation, with different distributions selectively influencing leading-edge shear-layer dynamics versus trailing-edge wake formation. A small number of studies on insect-inspired wings have begun to address spatially graded flexibility, identifying an optimal flexural rigidity distribution that maximises lift production, sensitive to the ratio of flapping and natural frequencies \citep{reade2022investigation, yang2026aerodynamic}, but these investigations are limited to harmonic hovering kinematics and idealised planforms without physiologically motivated camber or thickness. The present work addresses this gap by investigating two biologically motivated wing sections with a stiff leading-edge segment and a compliant trailing-edge region, and comparing their aerodynamic behaviour with that of a conventional NACA0012 foil under identical kinematic conditions.

Simulating two-way coupled fluid–structure interaction for geometries with physiologically motivated thickness and camber distributions introduces considerable computational challenges compared to the idealised flat-plate and symmetric NACA configurations that dominate the flexible-aerofoil literature. A pronounced thickness taper, with a thick leading-edge section transitioning to a thin trailing-edge, as observed in avian primary feathers and their engineered analogues, produces a highly non-uniform near-wall cell distribution in the body-fitted mesh considered in the Arbitrary Lagrangian-Eulerian approach. When large-amplitude passive deformation is imposed on such a geometry during plunging, the thin trailing-edge region undergoes large rotations and displacements that severely stress the adjacent mesh cells, generating high skewness and non-orthogonality that degrade the accuracy of the Navier–Stokes solution. Existing computational investigations of bio-inspired wings have either retained the structural complexity of the geometry, while considering the wing rigid \citep{harvey}, or retained passive compliance while reducing the geometry to a symmetric, uniform-thickness profile \citep{heathcote2007flexible, eldredge2010roles}. To the authors' knowledge, no prior study has simultaneously incorporated physiologically motivated camber, a chordwise thickness distribution, and a two-way coupled FSI framework under accelerated plunging conditions.

The present work addresses these gaps through a systematic numerical investigation of passively morphing aerofoils subjected to accelerated plunging in the low Reynolds number regime ($Re = 10^3$). The present FSI framework couples an incompressible Navier–Stokes solver with a geometrically nonlinear structural model, and is capable of accommodating high deformations of variable-thickness avian-inspired wing profiles. Three different geometries are considered: a symmetric NACA0012 and two bio-inspired profiles that incorporate physiologically motivated camber and thickness distributions derived from primary-feather cross-sections of peregrine falcon and barn owl wings (\cite{harvey}). Structural stiffness and acceleration parameter are varied independently, enabling a systematic investigation of acceleration as the governing parameter in passively flexible FSI, the effect of which is well-established for rigid aerofoils under transient kinematics \citep{pullin2004unsteady}, but not yet examined for deformable lifting surfaces. 
The results are expected to establish acceleration as an independent governing parameter in passive morphing aerofoil aerodynamics, provide mechanistic insight into the vortex-structure coupling that determines the limits of force augmentation under transient forcing, and offer a physically grounded basis for the design of passively adaptive lifting surfaces operating in the unsteady environments.

The primary objectives of this paper are as follows: (i) to investigate the effect of wing geometry on aerodynamic performance, comparing a conventional NACA0012 airfoil to bio-inspired wing sections of the peregrine falcon and barn owl undergoing accelerated plunging kinematics, (ii) to examine the influence of non-dimensional bending rigidity on the dynamic response of different wing sections under accelerated plunging, (iii) to compare the response dynamics of different proportion trailing-edge flexible wing sections, (iv) to investigate the FSI response of different wing sections subjected to various acceleration parameters (transition speeds), and (v) to quantify leading-edge and trailing-edge vortex dynamics and assess their impact on aerodynamic forces.

The remainder of this paper is organised as follows. Section~\ref{sec:methodology} describes the computational methodology, including the governing equations, solver configuration, and the verification and validation studies undertaken to establish confidence in the numerical framework. Sections~\ref{sec:flexibility} and~\ref{sec:flexibility-extent} present the aerodynamic and structural response of the three wing geometries as functions of the non-dimensional bending rigidity $K_B$ and the chordwise proportion of the trailing-edge flexible segment, respectively. Section~\ref{sec:acceleration} examines the effect of the transition speed parameter $a_s$ on the FSI response, with particular attention to the vortex dynamics and deformation-force coupling under varying kinematic forcing. Finally, Section~\ref{sec:conclusions} summarises the principal findings, discusses their implications for the design of bio-inspired passively adaptive lifting surfaces, and identifies directions for future work.

\section{Computational methodology}
\label{sec:methodology}

\subsection{Problem definition}

\begin{figure}
	\centering
	\includegraphics[width=1\columnwidth]{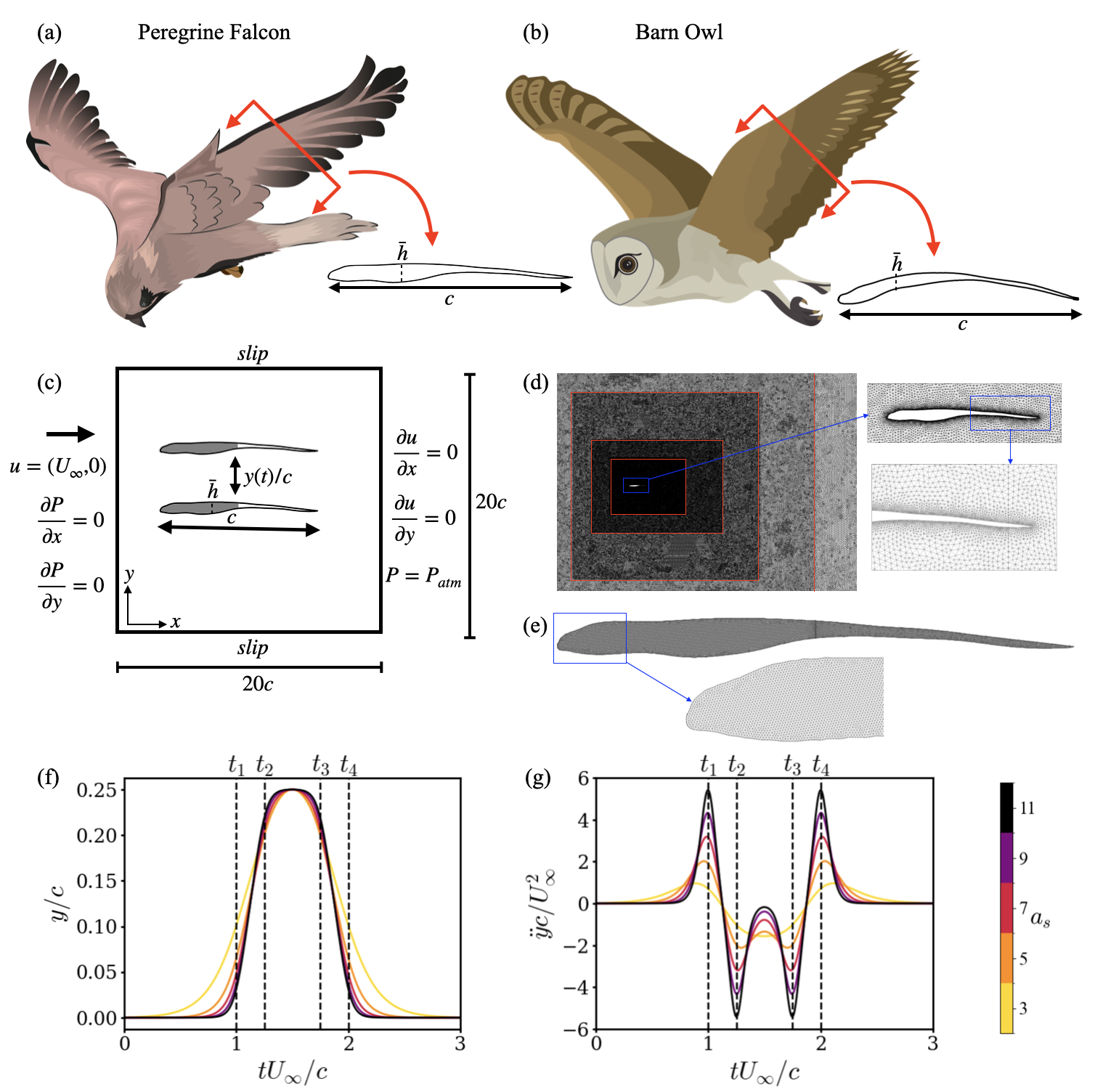}
	    \caption{Schematic representation of the bio-inspired wing sections of (a) peregrine falcon and (b) barn owl, investigated in this study; (c) the present computational domain with dimensions and boundary conditions (not drawn to scale); (d) the zoomed insets of the fluid mesh with multiple refinement zones indicated in red, (e) the representative solid mesh for the falcon wing section; the prescribed accelerated plunging kinematics (f) displacement, and (g) the corresponding accelerations across a range of transition-speed parameter ($a_s$) values investigated.}
    \label{figure: plunging problem set up}
\end{figure}

In this study, we investigate the FSI behaviour of three different morphing wing sections: a symmetric NACA0012 airfoil and two bio-inspired wing sections of peregrine falcon and barn owl, in the low Reynolds number regime ($Re = 10^3$); see the schematic representations in figures~\ref{figure: plunging problem set up}(a, b). The avian wing geometries for the peregrine falcon and barn owl are obtained from experimental measurements carried out by \cite{harvey}. The chord length ($c$) of each wing section is considered as the reference length scale, and the average thickness ($\Bar{h} = \frac{1}{c}\int_0^c h(x) \mathrm{d}x$) is considered for calculating the area moment of inertia. The bio-inspired wing sections are modelled with trailing-edge flexibility, where the wing segment from the mid-chord to the trailing-edge is flexible. 

The present two-dimensional computational domain dimensions and the associated boundary conditions are illustrated in figure~\ref{figure: plunging problem set up}(c). A $20c \times 20c$ square computational domain around the wing section is considered as the fluid domain; the inlet is positioned $5c$ upstream of the leading edge (0,0); the outlet is located $15c$ downstream; and the upper and lower walls are placed at a distance $10c$ from the leading-edge. At the inlet, a uniform free-stream velocity $U_\infty$ is applied, with zero gradient for pressure. At the outlet, the velocity satisfies a zero-gradient condition, while the pressure is fixed at atmospheric pressure. Slip boundary conditions are applied along the top and bottom boundaries to allow parallel flow to the walls. A moving-wall velocity condition and a zero-gradient pressure condition are applied to the wing section. The wing surface is defined as a fluid-structure interface to enable the transfer of aerodynamic forces and structural displacements to the solid wing mesh. 

An unstructured mesh is used due to the geometric complexity of avian-inspired wing sections. To improve resolution near the wing, the domain is divided into five sub-zones with varying mesh refinement as shown in figure~\ref{figure: plunging problem set up}(d). This strategy of using refinement zones ensured a smooth transition from coarser elements in the far-field to refined elements in the vicinity of the wing, thereby reducing computational cost compared to a fully refined mesh. The zoomed inset of the fluid mesh for the falcon wing section is shown in figure~\ref{figure: plunging problem set up}(d) for reference, as well as a close-up view of the mesh around the trailing-edge of the wing. Figure ~\ref{figure: plunging problem set up}(e) presents the representative solid mesh for the falcon wing section. It is worth noting that we use higher-order interpolation with radial-basis function (RBF) for data exchange at the fluid-solid interface; hence, a direct node-to-node mapping between the fluid and solid meshes is not required.

A transverse discrete temporal gust is mimicked by prescribing an accelerated plunging motion to the wing sections \citep{eldredge2009computational, wang2013low}. The plunging kinematics in terms of displacement are presented in figure~\ref{figure: plunging problem set up}(f), along with the corresponding accelerations in figure~\ref{figure: plunging problem set up}(g). For the plunging motion of the wing, the y-displacement is prescribed over time as:
\begin{equation}
   y(t) = y_0 \frac{G(t)}{G_{max}},
\label{eq: y(t)}
\end{equation}
where the maximum non-dimensional displacement, $y_0/c=0.25$ and $G(t)$ describes the acceleration phases of the plunging manoeuvre; see Eq.~\ref{eq: G(t)}. Here, $G_{max}$ denotes the maximum value of $G(t)$.
\begin{equation}
    G(t) = \ln \left[ \frac{\cosh(a_s U_{\infty} (t - t_1) / c) \cosh(a_s U_{\infty} (t - t_4) / c)}{\cosh(a_s U_{\infty} (t - t_2) / c) \cosh(a_s U_{\infty} (t - t_3) / c)} \right].
    \label{eq: G(t)}
\end{equation}
The plunging manoeuvre is divided into four key instants: $t_1 = c/U_\infty$ marks the onset of the ramp-up motion; $t_2 = t_1 + y_0/\dot{y_0}$ denotes its completion; $t_3 = t_2 + {\Delta}T_h$ indicates the start of the ramp-down and $t_4$ = $t_3$ + $y_0/\dot{y_0}$ denotes its completion; the total gust period considered in this study is $tU_\infty/c=3$. The time instances are marked in figures~\ref{figure: plunging problem set up}(f-g). The manoeuvre begins at $t_1$ = $c/U_\infty$, allowing sufficient time for boundary layer development before the onset of plunging motion. The hold time between the ramp-up and ramp-down phases, $t_2$ and $t_3$, is set to ${\Delta}T_h U_\infty/c=0.5$, and the rate of motion is set to $\dot{y_0}/U_\infty=1$. The transition rate between time intervals is controlled by the transition speed parameter $a_s$. Smaller values of $a_s$ represent a gradual transition, whereas larger values result in sharper, more abrupt transitions in motion, as shown in figure~\ref{figure: plunging problem set up}(f). In this study, five different transition speed parameters are investigated: $a_s=3, 5, 7, 9$ and $11$. The case with $a_s=3$ represents a gradual gust, whereas $a_s=11$ corresponds to a severe, sudden gust.

\subsection{Governing equations}

The laminar incompressible flow is governed by the Navier-Stokes (N-S) equations, given by:
\begin{equation}
    \nabla \cdot \mathbf{u} = 0, 
    \label{continuity}
\end{equation}

\begin{equation}
    \frac{\partial \mathbf{u}}{\partial t} +  (\mathbf{u}\cdot \nabla )\mathbf{u} = -\frac{1}{\rho_f}\nabla p + \nu \nabla^2 \mathbf{u},
    \label{NS}
\end{equation}

\noindent where \textbf{u} is the fluid velocity, $p$ is the fluid pressure, $\rho_f$ is the fluid density and $\nu$ is the Kinematic viscosity of the fluid. The N-S equations are directly solved using the finite-volume method without any turbulence modelling. The simulations are carried out using the open-source code {\tt OpenFOAM} \citep{weller1998tensorial,jasak2007openfoam}.

The nonlinear structural response of the wing sections is computed by solving the solid governing equations using the finite element method. The structural part is simulated using {\tt CalculiX} \citep{dhondt2017calculix}, an open-source finite element solver that can handle geometrically nonlinear problems involving large deformations. The material behaviour is described by a Neo-Hookean hyperelastic constitutive model, which is appropriate for the large-strain, large-deformation regime encountered at the wing trailing edge during accelerated plunging. The strain energy density function ($W$) is expressed in terms of the first invariant of the left Cauchy-Green deformation tensor and the Jacobian of the deformation gradient, ensuring material frame-indifference and incompressibility in the limiting case.

\begin{equation}
    W = C_{10}(\Bar{I_1} - 3) + \frac{1}{D_1}(J-1)^2,
    \label{equation: hyperelastic}
\end{equation}
where $D_1 = \frac{2}{\lambda}$, $C_{10} = \frac{\mu}{2}.$ $\Bar{I_1}$ is the first invariant of the deviatoric part of the right Cauchy-Green deformation tensor and
$J$ is the determinant of the deformation gradient. 
The shear modulus and the bulk modulus, $\mu$ and $\lambda$ respectively, are expressed in terms of Young's Modulus ($E$) and Poisson's ratio ($\nu_p$) as
$\mu = \frac{E}{2(1+\nu_p)}$ and
$\lambda = \frac{E}{3(1-2\nu_p)}$.

\begin{figure}
    \centering
    \includegraphics[width=0.8\textwidth]{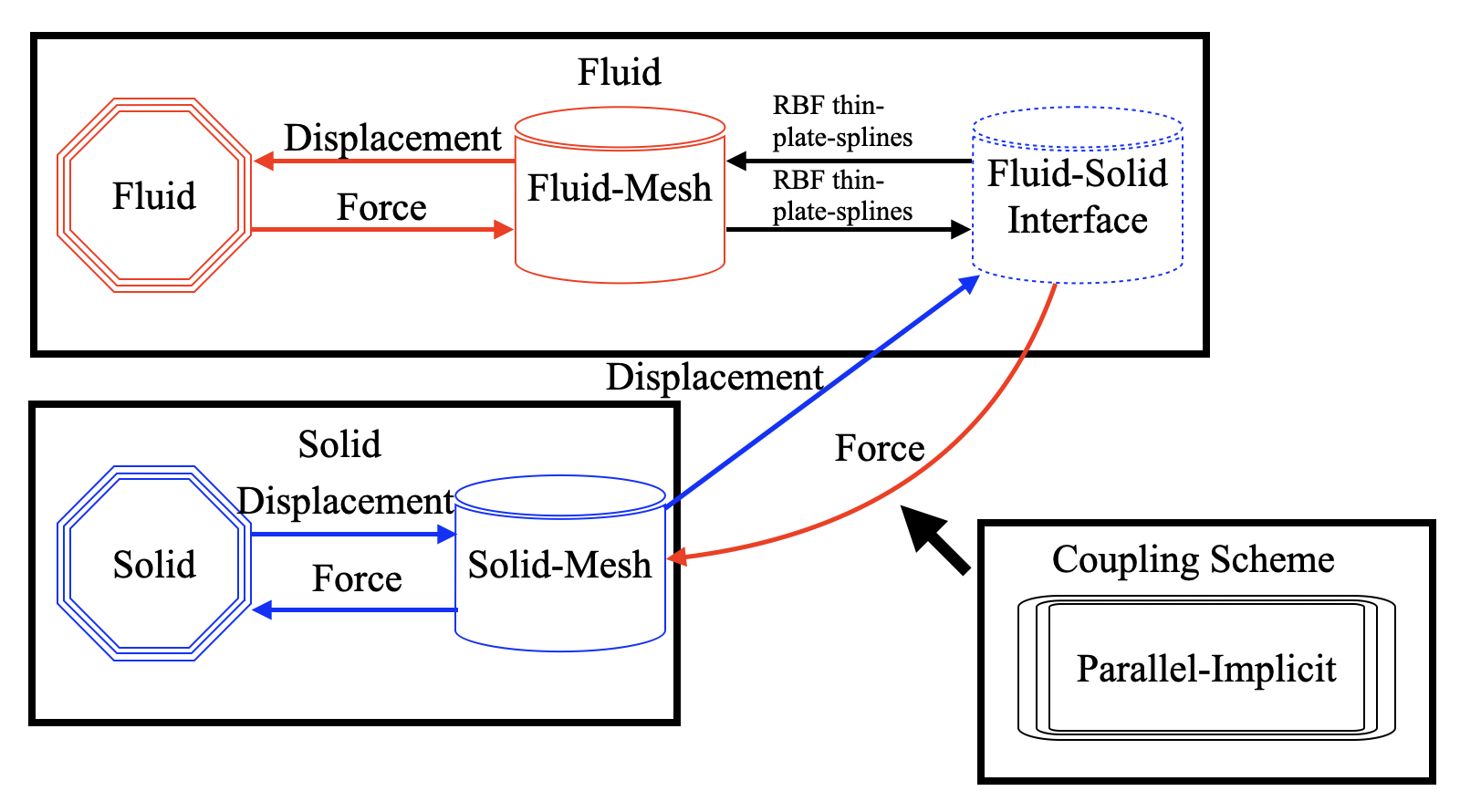}
    \caption{Schematic representation of the fluid-structure coupling framework.}
    \label{figure: preCICE}
\end{figure}

The two-way fluid-solid coupling is implemented using a partitioned strong-coupling approach through {\tt{preCICE}} \citep{chourdakis2023openfoam}. In this study, {\tt{preCICE}} couples {\tt{OpenFOAM}} and {\tt{CalculiX}} by facilitating the data exchange of forces from the fluid domain and displacements from the solid domain at the fluid-solid interface nodes, as shown visually in figure~\ref{figure: preCICE}. To ensure the accuracy of the FSI coupling, two interface conditions, the kinematic and dynamic coupling conditions, are satisfied \citep{shamanskiy2021mesh}. The kinematic condition enforces continuity of displacement and velocity across the fluid-solid interface, while the dynamic condition ensures equilibrium of forces between the two domains. The kinematic coupling conditions are expressed as
\begin{equation}
    x_f = w_s,
\end{equation}

\begin{equation}
    v_f = \frac{\partial w_s}{\partial t},
\end{equation}
\noindent where $x_f$ denotes the fluid interface position , $w_s$ the structural displacement, $v_f$ the fluid velocity at the interface, and $\frac{\partial w_s}{\partial t}$ the structural velocity. The dynamic coupling condition guarantees that the force or stress is balanced at the fluid-structure interface:
\begin{equation}
    \sigma_f \cdot n_f = -\sigma_s \cdot n_s,
    \label{eq:example}
\end{equation}
where $\sigma_f$ and $\sigma_s$ are the fluid and structural stress tensors, and $n_f$ and $n_s$ are the corresponding outward unit normal vectors.

Two key non-dimensional parameters are considered in this study, the non-dimensional bending rigidity ($K_B$) and the density ratio ($\rho^*$), defined as:

\begin{equation}
\label{mass-ratio_and_rigity}
K_B=\frac{EI}{\rho_f U_\infty^2 c^3}, \hspace{0.2cm} \mathrm{and} \hspace{0.2cm}
\rho^*=\frac{\rho_s}{\rho_f},\end{equation}

\noindent where $\rho_s$ is the solid density, and $I=\frac{b{\bar{h}^3}}{12}$ is the area moment of inertia. In the present study, we investigate the effect of flexibility by varying $K_B$ ($=5, 7.5, 10, 12.5, 15, 17.5, 20$ and $100$), while the $\rho^*$ value is kept fixed at $100$.

\subsection{Solver setup}

\noindent Flow simulations are performed using the {\tt{pimpleFoam}} solver, a transient and incompressible flow solver in {\tt{OpenFOAM}} based on the PIMPLE algorithm, a hybrid of the Pressure-Implicit with Splitting of Operators (PISO) and Semi-Implicit Method for Pressure-Linked Equations (SIMPLE) algorithms. To accommodate mesh deformation arising from both prescribed motion and flexible-wing deformation, a dynamic mesh-morphing strategy is employed. The {\tt{displacementLaplacian}} solver is used with quadratic inverse-distance diffusion to incorporate the mesh motion. The gradient and divergence terms in the governing equations are discretised using the Gauss linear scheme, while temporal discretisation is carried out using a second-order backward difference scheme. The Laplacian term is discretised with the Gauss linear corrected scheme. 
The pressure field is solved using the preconditioned conjugate gradient (PCG) solver and a diagonal-based incomplete Cholesky (DIC) preconditioner, with solver tolerances set to an absolute tolerance of $1\times10^{-8}$ and a relative tolerance of $1\times10^{-3}$. The velocity field and cell displacement are solved using the {\tt{smoothSolver}} in combination with a symmetric Gauss-Seidel smoother, with absolute and relative tolerance of $1\times10^{-6}$ and $1\times10^{-4}$, respectively. One non-orthogonal corrector loop is applied, and four corrector steps are performed per time step. A time step size of $\Delta t= 5 \times 10^{-4}$ is selected based on a time-independence study (discussed in the next subsection).

The geometrically nonlinear structural response of the flexible trailing-edge segment, involving significant deformation, is computed using the Neo-Hookean formulation in {\tt CalculiX} via the \texttt{*HYPERELASTIC} keyword. The structural response under dynamic loading is computed in {\tt{CalculiX}} via the \texttt{*DYNAMIC} keyword, employing direct time integration of the equations of motion. The \texttt{DIRECT} option is adopted, ensuring that the defined initial time increment remains constant. Time integration is performed using the generalised-alpha method \citep{miranda1989improved}, in which the numerical dissipation is controlled by the parameter $\alpha_d$, within the range $-1/3 \le \alpha_d \le 0$. In this study, $\alpha_d=-0.3$ is used, corresponding to near-maximum numerical dissipation \citep{dhondt2017calculix}. The implicit form of the algorithm is considered, which provides unconditional stability. At each time increment, the resulting system of linear equations is solved using the Spooles solver. The time step size in the solid solver is chosen as $\Delta t= 5 \times 10^{-4}$, consistent with the fluid time step size.

The two-way fluid-solid coupling is achieved using the partitioned strong (implicit) coupling strategy. Unlike in the weak-coupling strategy, this approach performs multiple sub-iterations within each coupling time step to satisfy the kinematic and dynamic interface conditions. For problems involving large structural deformations, such as those considered in this study, strong coupling provides improved stability and accuracy, as error accumulation in explicit schemes can lead to significant interface inconsistencies. Data transfer across the non-matching fluid-solid interface is performed using a radial basis function (RBF) interpolation scheme with thin-plate splines to ensure accurate data mapping. A parallel-implicit coupling scheme is employed, in which both solvers advance at each time step while exchanging interface data iteratively. Convergence acceleration is achieved using the Interface Quasi-Newton Inverse Least Squares (IQN-ILS) scheme, which leverages data from previous coupling iterations to improve subsequent predictions. Solver communication is managed via sockets, enabling communication when running the simulation across multiple computational nodes. The coupling time step is kept consistent with that of the fluid and the solid solvers, $\Delta t=5 \times 10^{-4}$.

\subsection{Verification studies}
\begin{figure}
    \centering
    \includegraphics[width=1\textwidth]{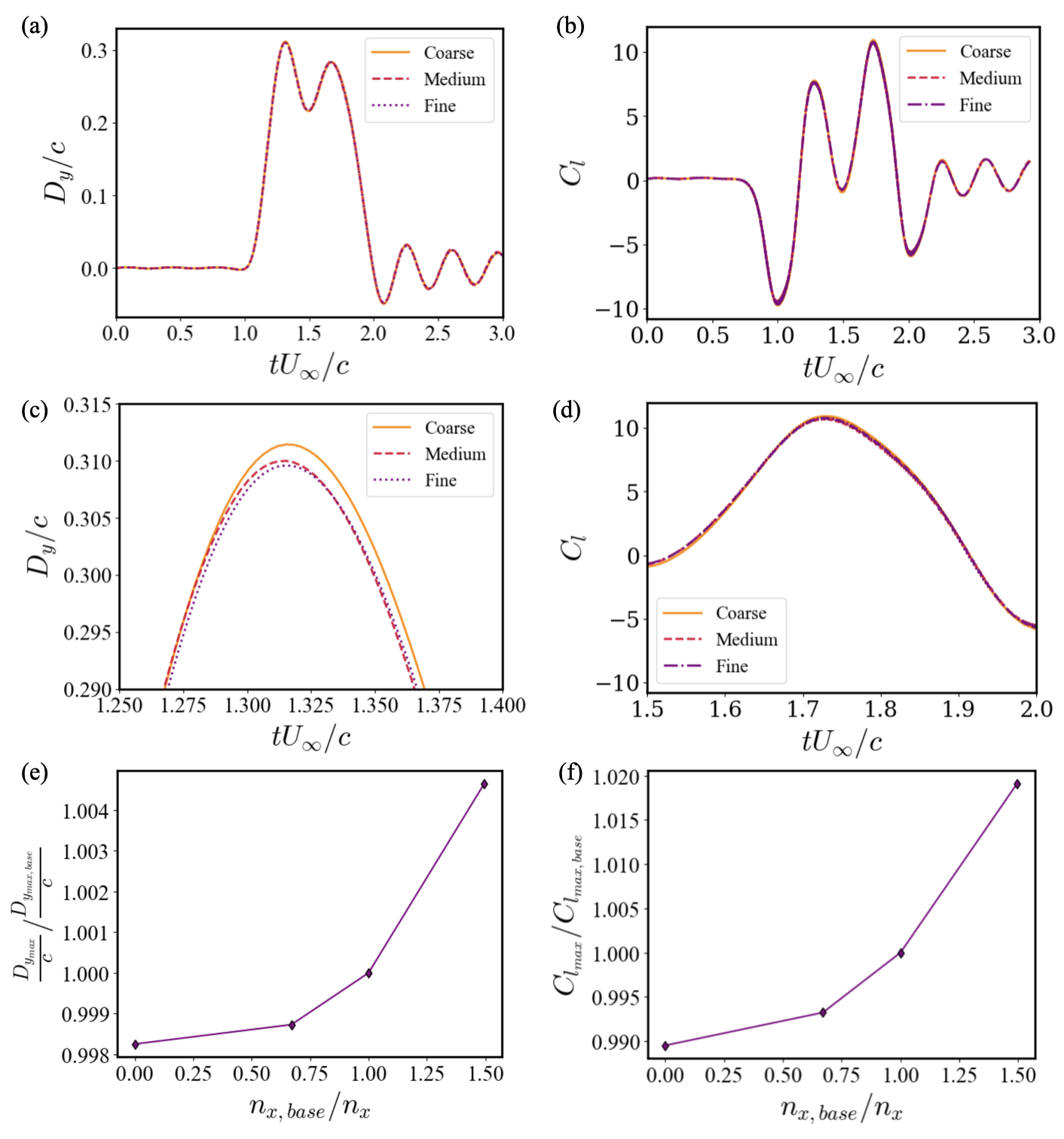}
    \caption{Grid independence study -- time histories of (a) normalised transverse displacement $D_y/c$ and (b) lift coefficient $C_l$ for the coarse, medium, and fine mesh configurations, with close-up views of the peak regions shown in (c) and (d), respectively. Panels (e) and (f) present the corresponding Richardson extrapolation curves for the peak values $D_{y_{\max}}/c$ and $C_{l_{\max}}$, respectively, demonstrating convergence of both quantities with mesh refinement.}
    \label{figure: verification study - meshes}
\end{figure}

A mesh independence study for the falcon wing section is conducted at $K_B=10$, $\rho^*=100$, and $a_s=11$ to choose an appropriate grid size, ensuring that the present simulation results are independent of the grid resolution. Three different meshes are tested, with cell counts of $5.86 \times 10^5$ (fine), $3.93 \times 10^5$ (medium), and $2.63 \times 10^5$ (coarse), respectively, and the corresponding non-dimensional trailing-edge displacement ($D_y/c$) and the coefficient of lift ($C_l$) time histories are compared; see figures~\ref{figure: verification study - meshes}(a, b). The zoomed insets are shown around their maximum values in figures~\ref {figure: verification study - meshes}(c, d). To further assess grid convergence, the mesh independence study is strengthened by the Richardson extrapolation \citep{celik2008procedure} using a refinement ratio of $r=1.5$. The extrapolated value represents the limiting value as the grid spacing tends to zero. The extrapolated results are shown in figures~\ref{figure: verification study - meshes}(e, f), where the medium and fine meshes are observed to converge towards the Richardson extrapolated value for both peak non-dimensional trailing-edge displacement ($D_{y_{max}}/c$) and the peak lift coefficient ($C_{l_{max}}$). The quantitative error analysis for $D_{y_{max}}/c$ and $C_{l_{max}}$ are summarised in Table.~\ref{table_mesh_time_step_Richardson_extrapolation}, where $\%$ relative errors are presented between the results obtained from extrapolated value and the medium mesh ($\boldsymbol{e_{extr}}$), fine and medium meshes ($\boldsymbol{e_{12}}$), and the medium and coarse meshes ($\boldsymbol{e_{23}}$). Based on this analysis, the medium mesh resolution is deemed sufficient for the present simulations, for which the maximum error relative to the extrapolated and fine-mesh results is approximately $1\%$, providing an appropriate balance between numerical accuracy and computational cost.

\begin{figure}
    \centering
    \includegraphics[width=1\textwidth]{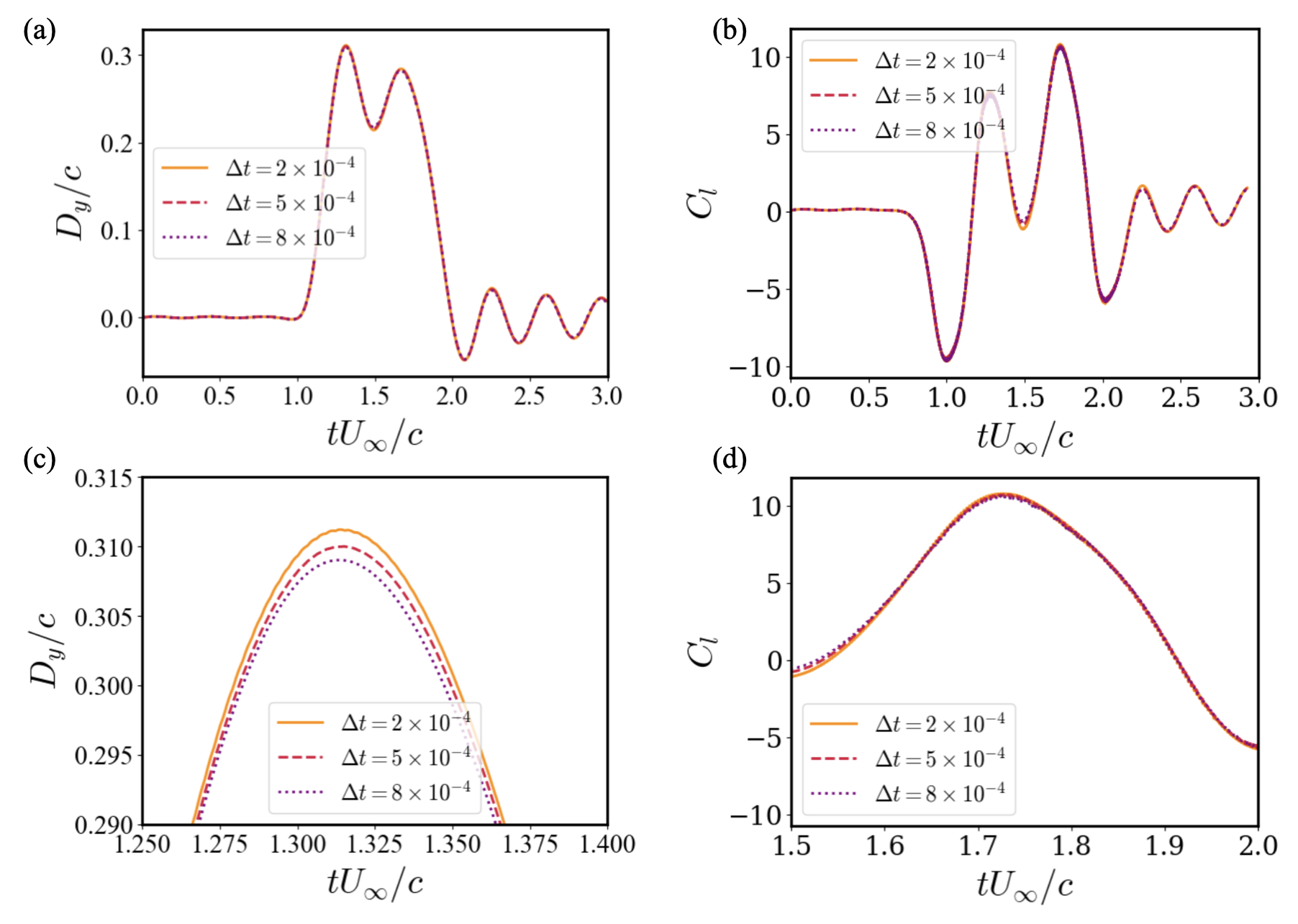}
    \caption{Time step independence study -- time histories of (a) normalised transverse displacement $D_y/c$ and (b) lift coefficient $C_l$ for the three time step sizes tested ($\Delta t =8 \times 10^{-4}$, $5 \times 10^{-4}$, and $2 \times 10^{-4}$), with close-up views of the peak regions shown in (c) and (d), respectively, demonstrating convergence of both quantities with temporal refinement.}
    \label{figure: verification study - time step}
\end{figure}

A time step sensitivity study is conducted using the selected medium mesh. Three time step sizes are tested: $\Delta t= 8 \times 10^{-4}$, $5 \times 10^{-4}$, and $2 \times 10^{-4}$. The temporal evolution of $D_y/c$ and $C_l$ for each time step is presented in figures~\ref{figure: verification study - time step}(a, b), along with close-up views around their maximum values in figures~\ref{figure: verification study - time step}(c, d). A comparison of $D_{y_{max}}/c$ and $C_{l_{max}}$ for the three time step sizes is provided in Table.~\ref{table_mesh_time_step_Richardson_extrapolation}, where the relative error between $\Delta t = 2 \times 10^{-4}$ and $5 \times 10^{-4}$ is less than $0.7\%$. Given the substantially higher computational cost associated with $\Delta t = 2 \times 10^{-4}$, a time step size of $\Delta t = 5 \times 10^{-4}$ is considered sufficient for all simulations. The chosen time-step size ensures that the maximum local Courant number, computed using the length of the smallest cell in the grid, remains less than 1 throughout the simulation.

\begin{table}
  \begin{center}
\def~{\hphantom{0}}
  \begin{tabular}{lccccccc}
      \textbf{Parameter}
      & \multicolumn{3}{c}{}
      & \textbf{Extrapolated}
      & $\boldsymbol{e_{12}}\ \boldsymbol{[\%]}$

      & $\boldsymbol{e_{23}}\ \boldsymbol{[\%]}$

      & $\boldsymbol{e_{\mathrm{extr}}[\%]}$ \\[3pt]

      ~ & \textbf{Fine} &\textbf{ Medium}& \textbf{Coarse} & \textbf{solution} & ~ & ~ & ~ \\[3pt]

      $D_{y_{max}}/c$
        & 0.3096 & 0.3100  & 0.3114
        & 0.3095
        & 0.13  & 0.45 & 0.16 \\

      $C_{l_{max}}$
        & 10.6788 & 10.7517 & 10.9572
        & 10.6388
        & 0.68 & 1.91 & 1.05 \\[6pt]

      ~ & $\boldsymbol{\Delta t = 2 \times 10^{-4}\,\mathrm{s}}$
 & $\boldsymbol{\Delta t = 5 \times 10^{-4}\,\mathrm{s}}$
 & $\boldsymbol{\Delta t = 8 \times 10^{-4}\,\mathrm{s}}$

      & ~ & ~ & ~ & ~ \\[3pt]

      $D_{y_{max}}/c$
        & 0.3112 & 0.3100 & 0.3090
        & -
        & 0.39  & 0.32 &  - \\

      $C_{l_{max}}$
        &10.8238 & 10.7517  & 10.6667
        & - 
        & 0.67 & 0.79& - \\
  \end{tabular}
  \caption{Percentage relative errors between $D_{y_{max}}/c$ and $C_{l_{max}}$ results obtained from the three different meshes and time step sizes.}
\label{table_mesh_time_step_Richardson_extrapolation}
  \end{center}
\end{table}

\subsection{Solver validation}
\begin{figure}
    \centering
    \includegraphics[width=1\textwidth]{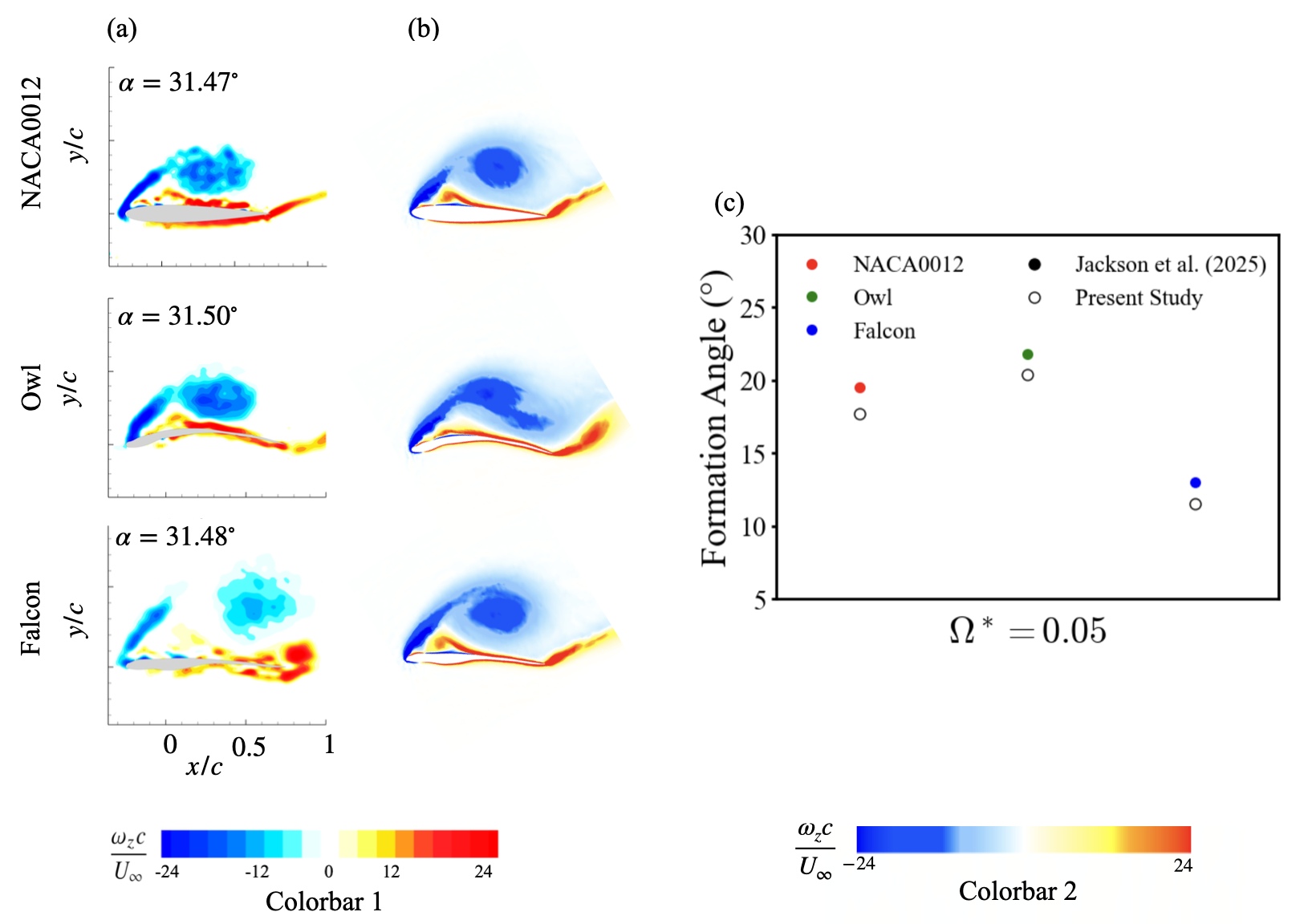}    
    \caption{Flow solver validation -- (a) vorticity contours around the NACA0012 foil, falcon, and owl wing sections reproduced from \citet{harvey} and (b) corresponding contours obtained from the present simulations, where Colorbar 1 and Colorbar 2 denote the colour scales associated with the reference and present results, respectively; (c) a comparison of the dynamic stall vortex formation angle for the three wing sections against the experimental measurements of \citet{harvey}, demonstrating close agreement across all geometries.}
    \label{figure: validation jackson}
\end{figure}

We first perform a flow-solver validation study for the three wing sections (NACA0012, falcon, and owl), assuming they are rigid, to compare the present flow-field results with those reported by \cite{harvey}. Each wing is subjected to a constant linear pitching motion, characterised by a constant non-dimensional pitch rate ($\Omega^* = \frac{{\dot{\alpha}c}}{{2U_{\infty}}} = 0.05$) at a chord-based Reynolds number of $Re_c=12,000$. 
The present results show a close agreement with the observations reported by \cite{harvey}, with comparable LEV locations and sizes; see figures~\ref{figure: validation jackson}(a, b). The formation angle is defined as the angle of attack at which the dynamic stall vortex first develops near the leading-edge of the wing. To determine this angle, we employ a method proposed by \cite{hrynuk2020effects} that identifies the initial onset of secondary vorticity. Specifically, the formation point corresponds to the initial sign change in vorticity (from negative to positive) near the leading-edge. The formation angles obtained for all three wing sections from the present simulations are compared against those reported by \cite{harvey}. Although the present formation angles are consistently lower by a small margin relative to the reference results, the differences are small across all cases, as shown in figure~\ref{figure: validation jackson}(c).

\begin{figure}
    \centering
    \includegraphics[width=1\textwidth]{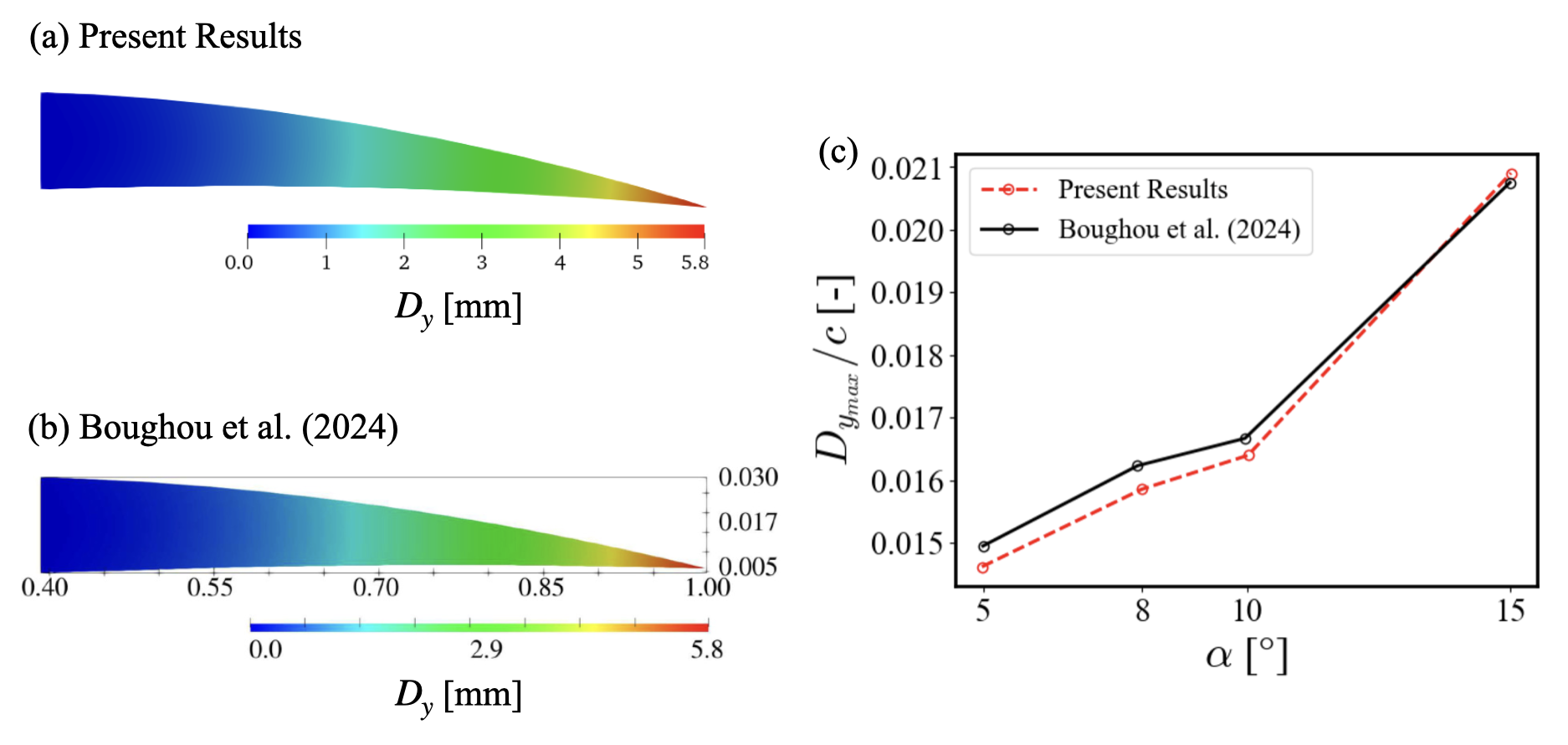}    
    \caption{FSI solver validation -- displacement contours of the flexible trailing-edge segment at $\alpha = 15^\circ$ obtained from (a) the present simulations and (b) the reference results of \citet{boughou2024investigation}, with the colourbar scale denoting the magnitude of transverse displacement $D_y$ in millimetres; (c) a comparison of the normalised maximum trailing-edge tip displacement $D_{y_{\max}}/c$ as a function of angle of attack $\alpha$ between the present results and those reported by \citet{boughou2024investigation}, demonstrating close agreement across the full range of angles of attack considered.}
    \label{figure: validation Boughou}
\end{figure}

Next, to validate the coupled FSI framework adopted in this study, the present results for the structural response of a stationary NACA6409 airfoil with $60\%$ flexible trailing-edge are compared with those reported by \cite{boughou2024investigation} for a range of angles-of-attack, $\alpha=5^\circ, 8^\circ, 10^\circ$ and $15^\circ$. The $k-\omega -\mathrm{SST}$ turbulence model is employed at $Re_c = 5 \times 10^5$. The fluid properties ($\rho_f=1.225$ $\mathrm{kg/m^3}$ and $\nu = 1.4607 \times 10^{-5} \ \mathrm{m^2/s}$) and the structural properties ($E=5.6$ MPa, $\nu_p=0.3$ and $\rho_s=1500$ $\mathrm{kg/m^3}$) are kept consistent with those reported in the reference study. In the present simulation, the structural response of the flexible segment is modelled using the Neo-Hookean constitutive law. Figures~\ref{figure: validation Boughou}(a, b) compare displacement contours of the flexible segment at $\alpha = 15 ^\circ$, with the present results shown in figure~\ref{figure: validation Boughou}(a) and those presented by \cite{boughou2024investigation} in figure~\ref{figure: validation Boughou}(b).
Both contours show similar levels of deformation, characterised by a progressive increase in displacement towards the trailing-edge tip, where the maximum deflection occurs. Furthermore, figure~\ref{figure: validation Boughou}(c) presents the $D_{y_{max}}/c$ values across the range of tested $\alpha$, compared with those reported by \cite{boughou2024investigation}. The present results demonstrate close agreement with the reference data and capture the same monotonic increase in $D_{y_{max}}/c$ with increasing $\alpha$. Minor discrepancies are observed at lower $\alpha$ in $D_{y_{max}}/c$, which can be attributed to differences in computational factors, such as mesh resolution, numerical uncertainty, or the absence of structural damping in the present model, which is considered in the reference study. The quantitative comparison, including the relative errors with respect to the reference study, is summarised in Table \ref{validation-relative-error}. The relative errors remain below 2.5\% for all tested cases, confirming the accuracy of the present computational methodology.


\begin{table}
  \begin{center}
\def~{\hphantom{0}}
 \begin{tabular}{ p{4 cm} p{2cm} p{2cm} p{2cm} p{2cm}}
      {$\alpha$}& {$5^\circ$} & {$8^\circ$} & {$10^\circ$} & {$15^\circ$} \\[3pt]
       {Relative error in $D_{y_{\max}}/c$ [\%]}  & 2.0743  &  2.4045 & 1.6454 & 0.6879\\
  \end{tabular}
\caption{Percentage relative error in the $D_{y_{\max}}/c$ values obtained from the present simulation and those presented in \citet{boughou2024investigation}.}
\label{validation-relative-error}
  \end{center}
\end{table}

\section{Effect of non-dimensional bending rigidity of flexible trailing-edge}
\label{sec:flexibility}

\begin{figure}
    \centering
    \includegraphics[width=1\textwidth]{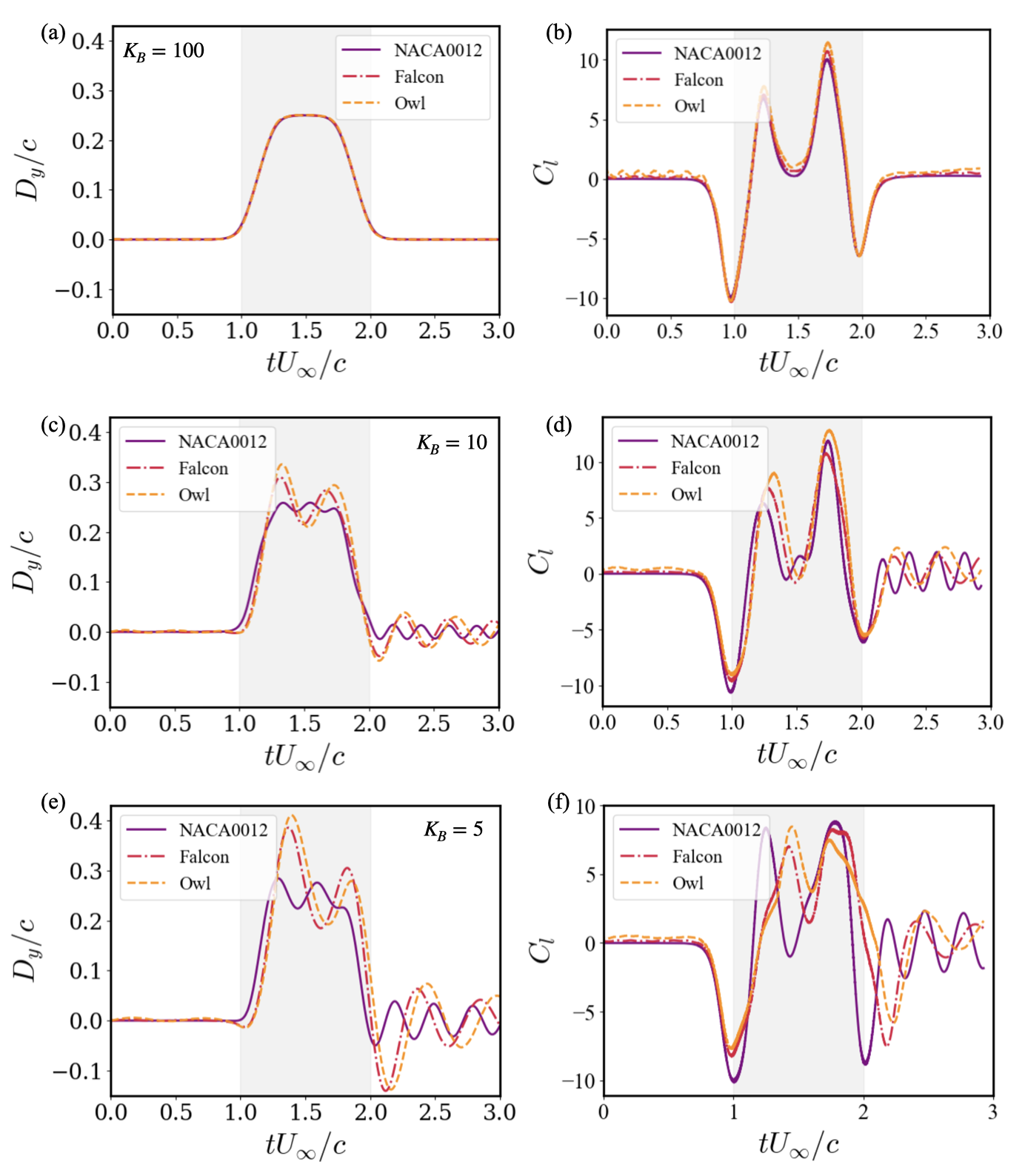}
    \caption{Time histories of normalised transverse displacement $D_y/c$ and lift coefficient $C_l$ for the NACA0012 foil, falcon, and owl wing sections at $a_s = 11$, comparing three bending stiffness values: $K_B = 100$ (a, b), $K_B = 10$ (c, d), and $K_B = 5$ (e, f). The shaded grey region denotes the duration of the accelerated plunging manoeuvre, corresponding to the interval $t_1$ to $t_4$ defined in figure~\ref{figure: plunging problem set up}(f).}
    \label{figure: plunging - flexibility}
\end{figure}

First, we investigate the effect of flexibility on the response dynamics and aerodynamic forces for the conventional NACA0012 foil and two different bio-inspired wing sections (peregrine falcon and barn owl) subjected to the most severe gust condition considered in this study, corresponding to $a_s=11$. Figures~\ref{figure: plunging - flexibility}(a, b) present the time histories of $D_y/c$ and $C_l$ for $K_B=100$, representative of a highly stiff wing with negligible passive deformation, when subjected to accelerated plunging. As expected, it is seen that the $D_y/c$ time histories (see figure~\ref{figure: plunging - flexibility}(a)) for all three wing sections closely follow the prescribed accelerated kinematics, shown in figure~\ref{figure: plunging problem set up}(f). Two distinct peaks are observed in the $C_l$ time history, corresponding to the upward and downward acceleration phases. A slight deviation in the magnitude of the peak $C_l$ values is seen for the three different wing sections at $K_B=100$; see figure~\ref{figure: plunging - flexibility}(b). Both the $C_l$ peaks attain their largest values for the owl wing section, followed by the falcon wing section, and then the conventional NACA0012 foil, underscoring the influence of wing geometry on lift generation even in the absence of significant structural deformation. 

Next, the response and lift characteristics are examined for more compliant configurations at lower values of $K_B$. Figures~\ref{figure: plunging - flexibility}(c, d) and \ref{figure: plunging - flexibility}(e, f) illustrate the transient wing response and aerodynamic lift for $K_B=10$ and $K_B=5$, respectively. The deformation response ($D_y/c$) for the bio-inspired wing sections exhibits a delayed onset relative to the kinematic acceleration, indicating a finite structural adaptation timescale, with the delay increasing as $K_B$ decreases. In contrast, for the NACA0012 foil, variations in $D_y/c$ and $C_l$ relative to the stiffer case ($K_B=100$), presented in figures~\ref{figure: plunging - flexibility}(a, b), are minimal across all $K_B$ values, indicating limited sensitivity to variations in chordwise flexibility. The owl wing section consistently shows the largest deformation amplitude, followed by the falcon wing section, while the NACA0012 foil exhibits the smallest deflection. This hierarchy directly reflects differences in effective compliance and in the chordwise distribution of stiffness. The lift response broadly follows the deformation behaviour, though with pronounced transient overshoots attributable to unsteady vortex dynamics. The prescribed kinematic acceleration induces a sharp increase in $C_l$ values that substantially exceed those typical of periodic kinematics. The large negative lift spike preceding the positive peak indicates rapid vortex formation accompanied by modulation of the leading-edge suction. Notably, the owl wing section generates the strongest lift amplification, suggesting that enhanced flexibility promotes transient vortex strengthening rather than damping. The phase difference between the deformation growth and the lift peaks indicates that elastic lag governs force production, rather than instantaneous kinematics alone.

\begin{figure}
    \centering
   \includegraphics[width=1\textwidth]{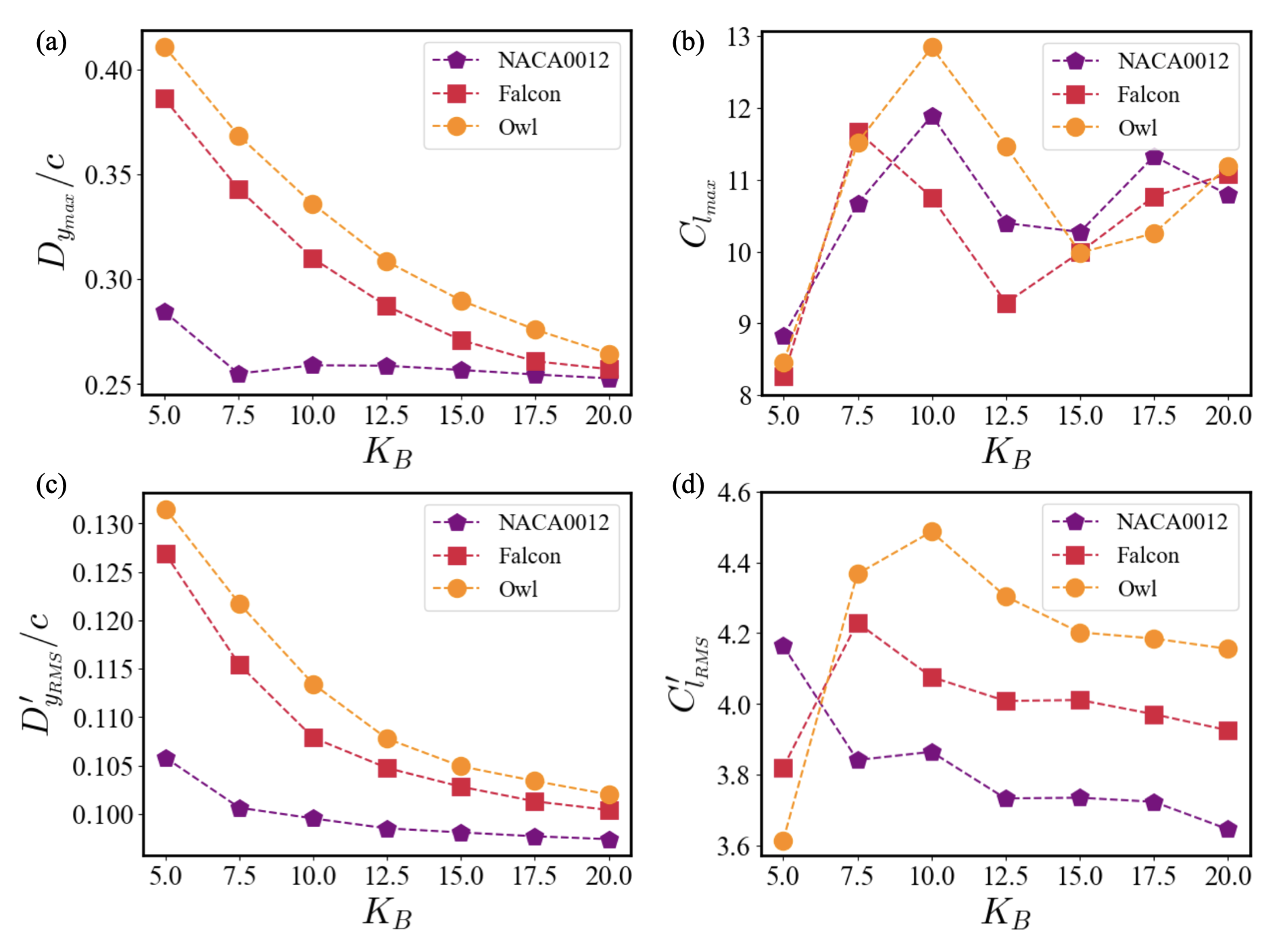}
    \caption{Variation of structural deformation and aerodynamic lift metrics with bending stiffness $K_B$ for the NACA0012 foil, falcon, and owl wing sections at $a_s = 11$: (a) peak normalised trailing-edge displacement $D_{y_{\max}}/c$, (b) peak lift coefficient $C_{l_{\max}}$, (c) root-mean-square of the mean-subtracted normalised trailing-edge displacement $D'_{y_{RMS}}/c$, and (d) root-mean-square of the mean-subtracted lift coefficient $C'_{l_{RMS}}$.}
    \label{figure: max values and RMS - flexibility}
\end{figure}

Figure~\ref{figure: max values and RMS - flexibility}(a) reveals a monotonic reduction in $D_{y_{max}}/c$ with increasing $K_B$, consistent with classical structural scaling. However, the sensitivity differs markedly across configurations. The owl wing section exhibits the steepest decay, suggesting a stronger dependence of deformation on flexibility. The NACA0012 foil exhibits minimal variation, indicating that the NACA0012 geometry is relatively insensitive to changes in bending stiffness across the range investigated. This behaviour indicates that geometry-dependent compliance distributions modulate acceleration-induced loading. Importantly, deformation does not collapse uniformly across designs, highlighting that stiffness alone is insufficient to predict morphing amplitude under transient forcing. Figure~\ref{figure: max values and RMS - flexibility}(b) presents the $C_{l_{max}}$ values across the $K_B$ values investigated. Unlike deformation, peak lift exhibits non-monotonic behaviour. Intermediate stiffness values yield the greatest lift amplification, indicating an optimal compliance regime. For the NACA0012 foil and owl wing sections, the maximum $C_l$ occurs for $K_B = 10$, whereas for the falcon wing section it occurs for $K_B = 7.5$, indicating that the optimal material properties are dependent on the wing geometry. Excessive flexibility reduces lift despite larger deformation, indicating that deformation magnitude alone does not guarantee force enhancement. This behaviour is consistent with a resonance-like mechanism, governed by deformation–vortex phase coupling. At moderate flexibility, the phase lag between structural deformation and aerodynamic loading is synchronised favourably with vortex growth, thereby maximising suction forces. At low $K_B$, excessive compliance disrupts this synchronisation, weakening vortex coherence and reducing lift. 

The root-mean-square (RMS) value of the non-dimensional trailing-edge displacement (${D}_{y_{RMS}}'/c$), shown in figure~\ref{figure: max values and RMS - flexibility}(c), exhibits the same monotonic trend, with the largest values observed for the owl wing section, followed by the falcon wing section, and the NACA0012 foil, which shows minimal variation across flexibilities. The RMS trends of $C_l$ shown in figure~\ref{figure: max values and RMS - flexibility}(d) reinforce this interpretation. The owl wing section exhibits larger fluctuations at lower $K_B$, reflecting stronger unsteady vortex activity. Increasing stiffness attenuates fluctuations, implying stabilisation of wake dynamics. The NACA0012 foil shows relatively weak sensitivity, consistent with its limited deformation response. These results suggest that flexibility primarily modulates variability in unsteady forces rather than mean loading, suggesting a link between structural compliance and wake stability.

\begin{figure}
    \centering
    \includegraphics[width=1\textwidth]{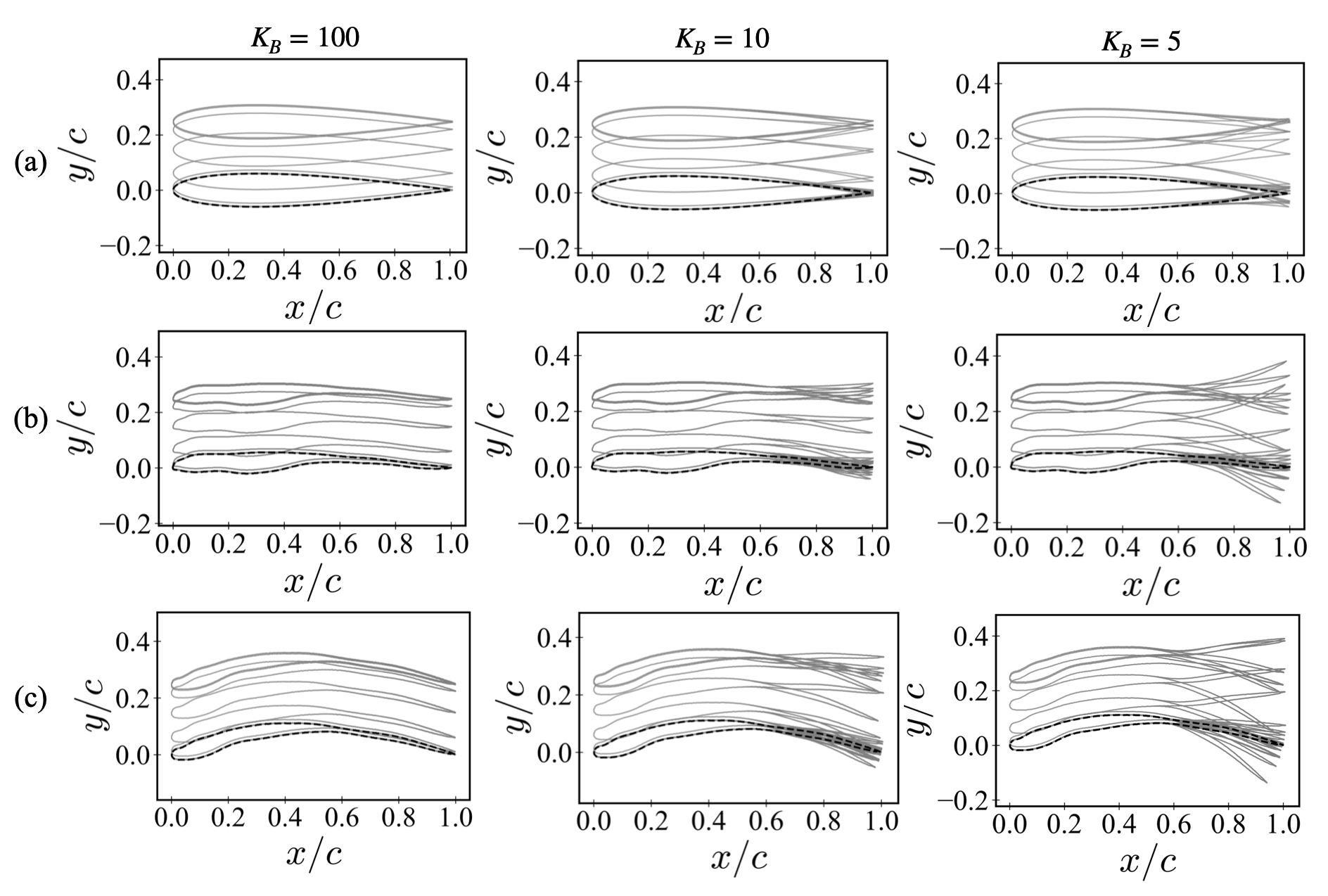}
    \caption{Displacement envelopes over the full gust manoeuvre for the (a) NACA0012 foil, (b) falcon, and (c) owl wing sections at $a_s = 11$, comparing three bending stiffness values: $K_B = 100$ (left column), $K_B = 10$ (centre column), and $K_B = 5$ (right column). Each grey line represents the instantaneous deformed wing profile at a discrete time instant, and the dashed black line denotes the undeformed reference geometry at the start.}
    \label{figure: displacement envelopes - flexibility}
\end{figure}

\begin{figure}
	\centering
	\includegraphics[width=1\columnwidth]{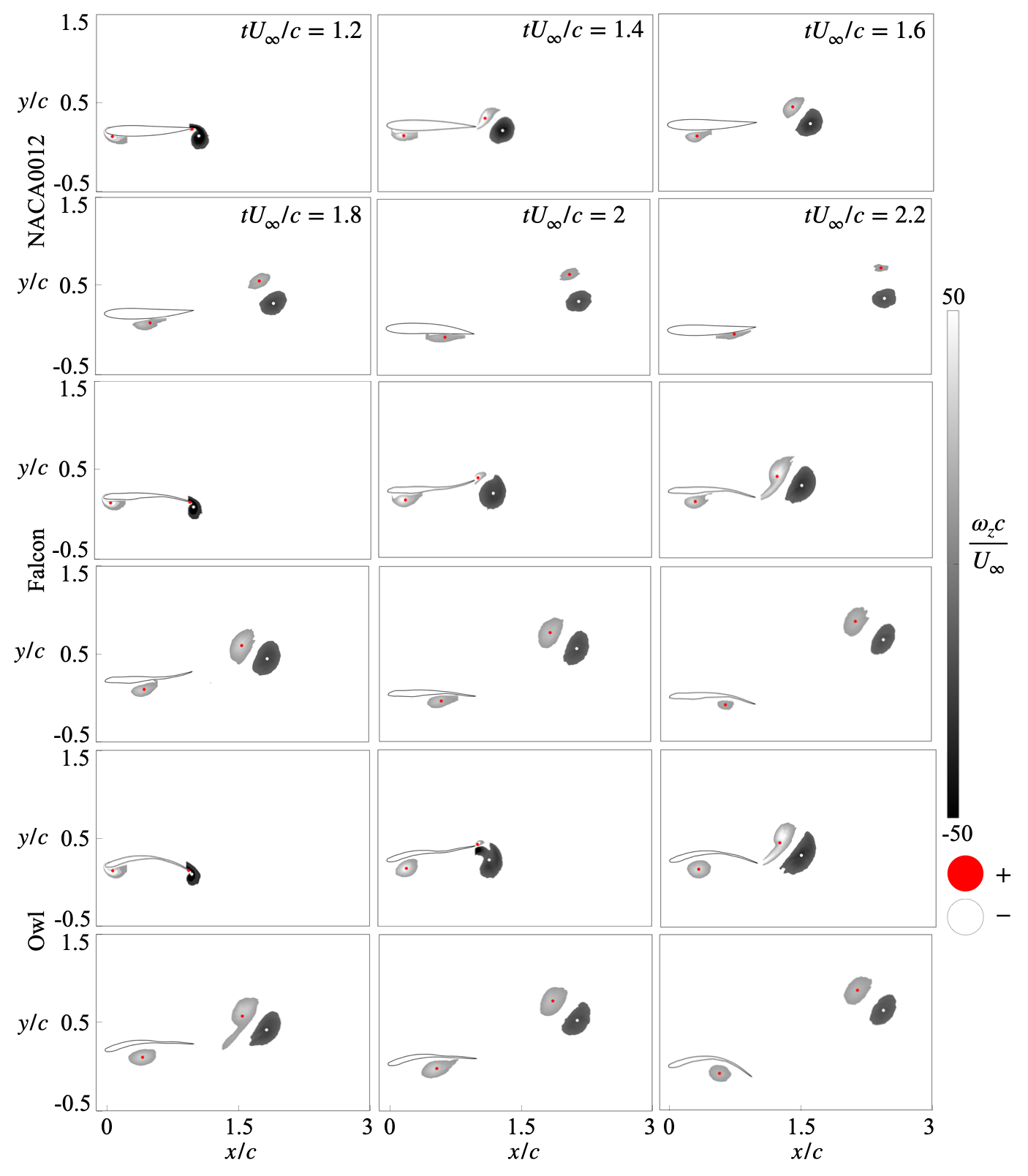}
	    \caption{Instantaneous vorticity contours $\omega_z U_\infty/c$ for the NACA0012 foil (rows 1--2), falcon (rows 3--4), and owl (rows 5--6) wing sections at $K_B = 5$ and $a_s = 11$, shown at six representative time instants: $tU_\infty/c = 1.2$, $1.4$, $1.6$, $1.8$, $2.0$, and $2.2$. The red and white filled circles denote the centroids of the positive and negative vortices, respectively. The colour bar scale is set at $-50 \le \omega_z U_\infty/c \le +50$.}
    \label{figure: fig vorticity contour}
\end{figure}

Figure~\ref{figure: displacement envelopes - flexibility} presents the deformation envelopes for the three different wing sections: the (a) NACA0012, (b) falcon, and (c) owl, for $K_B = 100$, $10$, and $5$ (refer to Supplementary video 1). The greatest deformation occurs in the most flexible ($K_B=5$) cases, whereas the least deformation is observed in the stiff ($K_B=100$) cases; the intermediate ($K_B=10$) cases exhibit moderate deformation. For $K_B=5$, the symmetric NACA0012 foil exhibits small-amplitude, oscillating trailing-edge deformation. The falcon wing section shows pronounced trailing-edge bending, and the owl wing section displays large-amplitude, spatially distributed curvature, indicating that multiple structural modes contribute significantly to the deformation response. This is consistent with the time histories presented in figure~\ref{figure: plunging - flexibility}, which show that even at low $K_B$, the trailing-edge deformation of the NACA0012 foil remains minimal relative to the bio-inspired geometries. The deformation shapes reveal that flexibility dynamically redistributes effective camber, thereby altering instantaneous aerodynamic characteristics. The spatially distributed curvature, observed at low $K_B$, is indicative of nonlinear coupling between the aerodynamic loading and the structural response, which is amplified under transient acceleration.

\begin{figure}
    \centering
    \includegraphics[width=1\textwidth]{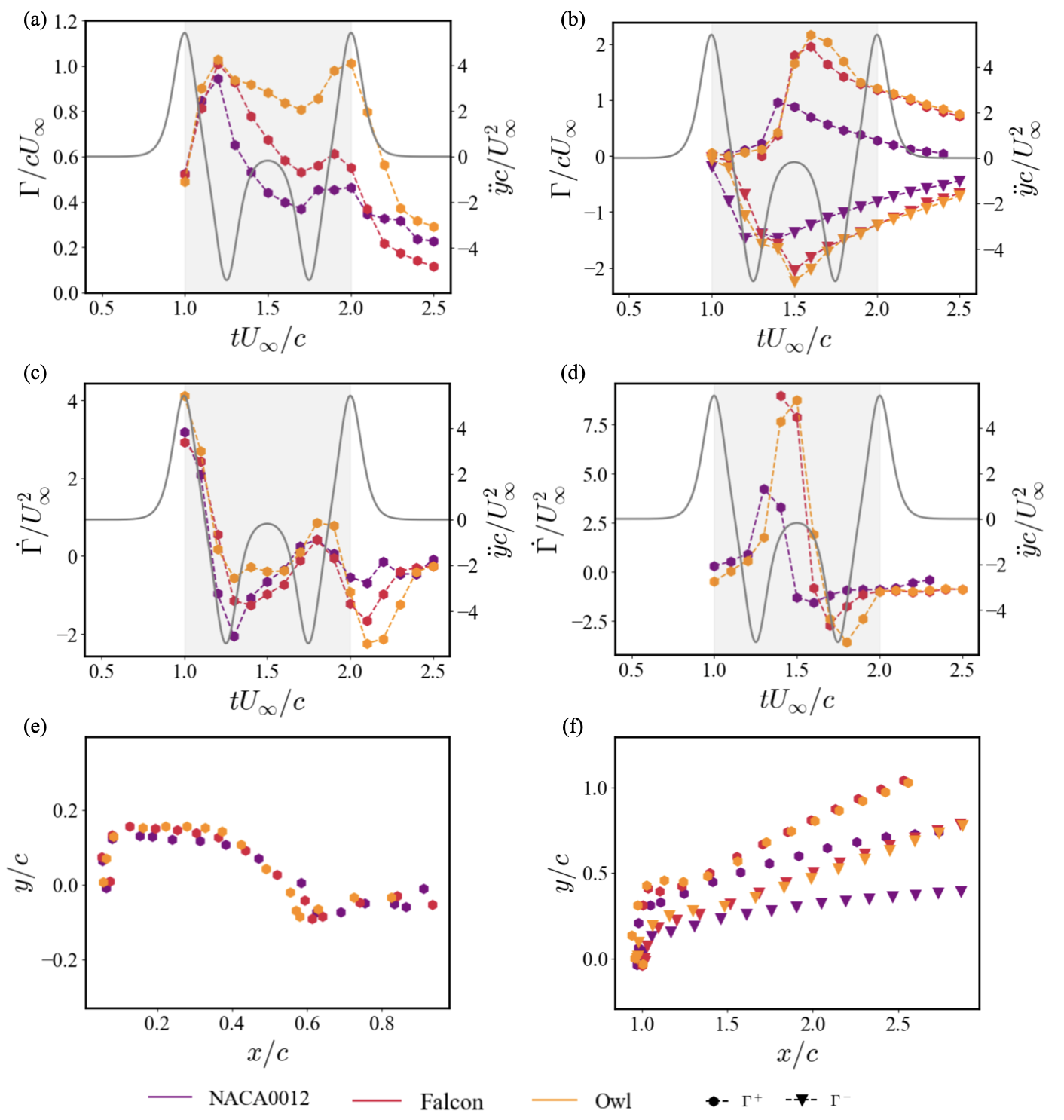}
    \caption{Vortex circulation analysis for the NACA0012 foil, falcon, and owl wing sections at $K_B = 5$ and $a_s = 11$, over the time window $tU_\infty/c = 1$ to $2.5$. Sub-figures (a) and (b) show the time histories of normalised circulation $\Gamma/cU_\infty$ for the leading-edge vortex (LEV) and trailing-edge vortex (TEV), respectively, with positive ($\Gamma^+$) and negative ($\Gamma^-$) contributions distinguished by filled circle and inverted triangle markers. Sub-figures (c) and (d) present the corresponding rates of circulation growth $\dot{\Gamma}/U_\infty^2$ for the LEV and TEV, respectively. The normalised wing acceleration $\ddot{y}c/U_\infty^2$ is overlaid as a solid grey line (a-d) to illustrate the phase relationship between kinematic forcing and vortex development. Sub-figures (e) and (f) show the trajectories of the LEV and TEV centroids, respectively, over the same time window, expressed in the wing reference frame.}
    \label{figure: LEV TEV circulation - flexibility}
\end{figure}

Figure~\ref{figure: fig vorticity contour} presents the vorticity contours of the primary near-field wake structures -- the LEV and TEVs -- for the three wing sections at representative time instants, $tU_\infty/c = 1.2$, $1.4$, $1.6$, $1.8$, $2.0$, and $2.2$ (see also Supplementary video 2). Over this interval, the positive LEV grows and convects along the lower surface of the wing sections, eventually forming a vortex couple with the TEV. The vortex couple sheds and convects downstream with its self-induced velocity, undergoing deflection. It progressively weakens, and its area decreases as viscous diffusion dissipates its strength. Although additional TEV structures are present further downstream, only the primary vortex couple shed from the trailing-edge is shown, which corresponds to the circulation analysis, presented in figure~\ref{figure: LEV TEV circulation - flexibility}. The first peak in $C_l$ occurs at approximately $tU_\infty/c = 1.2$ for the NACA0012 foil and at approximately $tU_\infty/c = 1.4$ for the bio-inspired wing sections, at which instants all three profiles exhibit upward trailing-edge deflection. Conversely, at the $C_l$ troughs, occurring at approximately $tU_\infty/c = 2.0$ for the NACA0012 foil and $tU_\infty/c = 2.2$ for the bio-inspired wing sections, all profiles display downward trailing-edge deflection, consistent with the phase lag between structural deformation and aerodynamic forcing identified in the time histories. The key qualitative distinction between the three geometries lies in the coherence and strength of the vortical structures. For the NACA0012 foil, the LEV remains predominantly attached to the wing surface throughout the interval shown, and the TEVs are comparatively weak, dissipating rapidly; this is especially apparent at $tU_\infty/c = 2.2$, where the positive TEV is substantially diminished. The bio-inspired wing sections, and in particular the owl wing section, generate a markedly more coherent and stronger LEV, together with more persistent TEVs, reflecting the enhanced compliance and cambered geometry of these profiles. The TEV convection trajectories also differ between the NACA0012 foil and the bio-inspired wing sections, with the bio-inspired geometries exhibiting greater lateral displacement of the vortex centroids as the structures advect into the wake.

Figure~\ref{figure: LEV TEV circulation - flexibility} illustrates the sequential evolution of the LEV and TEV during the accelerated plunging cycle ($tU_\infty/c=1$ to $2.5$) for the three foils for $K_B=5$ and $a_s=11$. The normalised circulation values ($\Gamma/cU_\infty$) are presented in figures~\ref{figure: LEV TEV circulation - flexibility}(a, b) for the LEVs and TEVs, respectively, while figures~\ref{figure: LEV TEV circulation - flexibility}(c, d) show the corresponding rate of change in circulation with time ($\dot{\Gamma}/U_\infty^2$). The normalised wing acceleration $\ddot{y}c/U_\infty^2$ for $a_s=11$ is overlaid in grey. $\Gamma$ is calculated as the area integral of the vorticity of each cell within the detected vortex boundary using a threshold criterion of vorticity isosurface ($-10 \le \omega_z \le 10)$, as given by $\Gamma = \iint_S \omega_z \,dS$. Here, $\omega_z$ denotes the vorticity about the z-axis, $S$ is the area of the identified vortex structure. The vortex centroids are calculated as $(x_c, y_c) = (\frac{\iint_S \omega_z x dS}{\iint_S \omega_z dS}, \frac{\iint_S \omega_z y dS}{\iint_S \omega_z dS})$. The initial acceleration phase promotes rapid LEV formation, followed by the development of a coherent TEV as vorticity generated at the trailing-edge rolls into the near wake. As the cycle progresses, the LEV convects downstream and interacts with the nascent TEV, forming a vortex couple. This coupled structure subsequently undergoes alternating upward and downward deflections, reflecting a periodic reorganisation of the wake momentum. The alternating deflection of the vortex couple indicates a dynamically asymmetric wake response, governed by the phase relationships among LEV convection, TEV growth, and foil deformation. Acceleration enhances the strength and proximity of successive vortices, promoting interactions between the near-field wake structures. These observations confirm that transient kinematics fundamentally modify vortex topology, with LEV–TEV coupling emerging as the dominant mechanism underlying wake deflection and unsteady force generation.

Across all three wing sections, $\Gamma_{LEV}$ rises sharply immediately after the onset of acceleration, indicating that non-uniform kinematics generates leading-edge vorticity at a rate exceeding that of its convection downstream. The subsequent divergence of trends between the wing sections can be interpreted in terms of how curvature and thickness redistribute the local effective incidence and the leading-edge pressure gradient during the accelerating phase. The NACA0012 foil shows a rapid early increase followed by a comparatively strong decay. This can be attributed to the fact that acceleration initially triggers a strong LEV roll-up, but once the shear layer detaches and the LEV convects, circulation decays as feeding weakens and the LEV loses coherence. The falcon wing section exhibits a similar rise but sustains circulation longer than the NACA0012 foil. We hypothesise that moderate curvature produces a more favourable near-LE pressure recovery and maintains LE shear-layer attachment for longer during the acceleration window, thereby continuing to feed the LEV and keeping it coherent before shedding. The owl wing section (largest curvature and thickness distribution) shows the highest sustained $\Gamma_{LEV}$ over the accelerated portion of the motion. This indicates that its curvature/thickness distribution promotes stronger suction-side shear-layer vorticity feeding and LEV retention, delaying convective shedding. Physically, a more curved mean line increases the instantaneous effective camber under plunge, elevating the leading-edge suction peak during acceleration and enhancing shear-layer roll-up. A thicker leading-edge radius can also shift the separation point and stabilise a larger, more coherent LEV by moderating locally adverse pressure gradients, particularly under rapidly increasing effective angle of attack.

A pronounced change in $\Gamma_{TEV}$ is observed for all three wing sections around the same time interval in agreement with the canonical sequence: rapid LEV growth, strong induced velocities and bound-circulation variation, enhanced vorticity shedding at the TE. However, the contrast between wing sections is sharper here because curvature and thickness strongly influence the phase relationship between LEV evolution and trailing-edge shedding. The NACA0012 foil tends to exhibit an earlier, relatively weaker TEV response, suggesting that when the LEV sheds relatively early, the wake transitions more quickly to a convective regime and trailing-edge shedding becomes less extreme. The symmetric thickness distribution yields less pronounced modulation of the camber-induced bound circulation, limiting TEV amplification even under acceleration. The falcon wing section exhibits a stronger TEV excursion, evidenced by a more pronounced time-rate-of-change of bound circulation under acceleration. Moderate curvature can increase the amplitude of circulation modulation along the chord, strengthening the trailing-edge shear layer and producing a more energetic TEV during the wake reorganisation. The owl wing section shows the strongest and most persistent TEV circulation. This is consistent with a mechanism where a retained, stronger LEV induces larger near-wake velocities and maintains higher shear at the trailing-edge, while the curvature/thickness distribution sustains a larger bound-circulation swing during acceleration. In other words, the owl profile appears to convert acceleration-driven LEV growth more efficiently into trailing-edge shedding and wake momentum redistribution. Acceleration increases $|d\Gamma/dt|$ (both bound and vortex circulation), and the bio-inspired curvature/thickness amplifies the conversion of this unsteady bound-circulation change into TEV strength. This makes $\Gamma_{TEV}$ more sensitive than $\Gamma_{LEV}$ to geometric differences, because it integrates both leading-edge dynamics and chordwise circulation redistribution. The locations of the LEV and TEV centroids are shown in figures~\ref{figure: LEV TEV circulation - flexibility}(e, f) over the same time interval. The LEV centroid trajectories remain similar over time for all three wing sections. In contrast, differences are observed in the TEV centroid trajectories; the falcon and owl wing sections follow comparable paths that propagate further downstream and at steeper upward angles than the NACA0012 foil trajectory, which travels along a relatively straighter path. 

In summary, acceleration initiates LEV growth in all cases, but increasing curvature/thickness modulation shifts the response from a short-lived LEV of limited coherence with weaker trailing-edge processing (NACA0012 foil) to a sustained LEV that drives stronger TEV circulation through enhanced chordwise circulation modulation (falcon/owl), with the owl case showing the most efficient conversion of kinematic acceleration into vortex amplification.

\section{Effect of flexibility extent for the compliant trailing-edge}
\label{sec:flexibility-extent}
\begin{figure}
    \centering
    \includegraphics[width=1\textwidth]{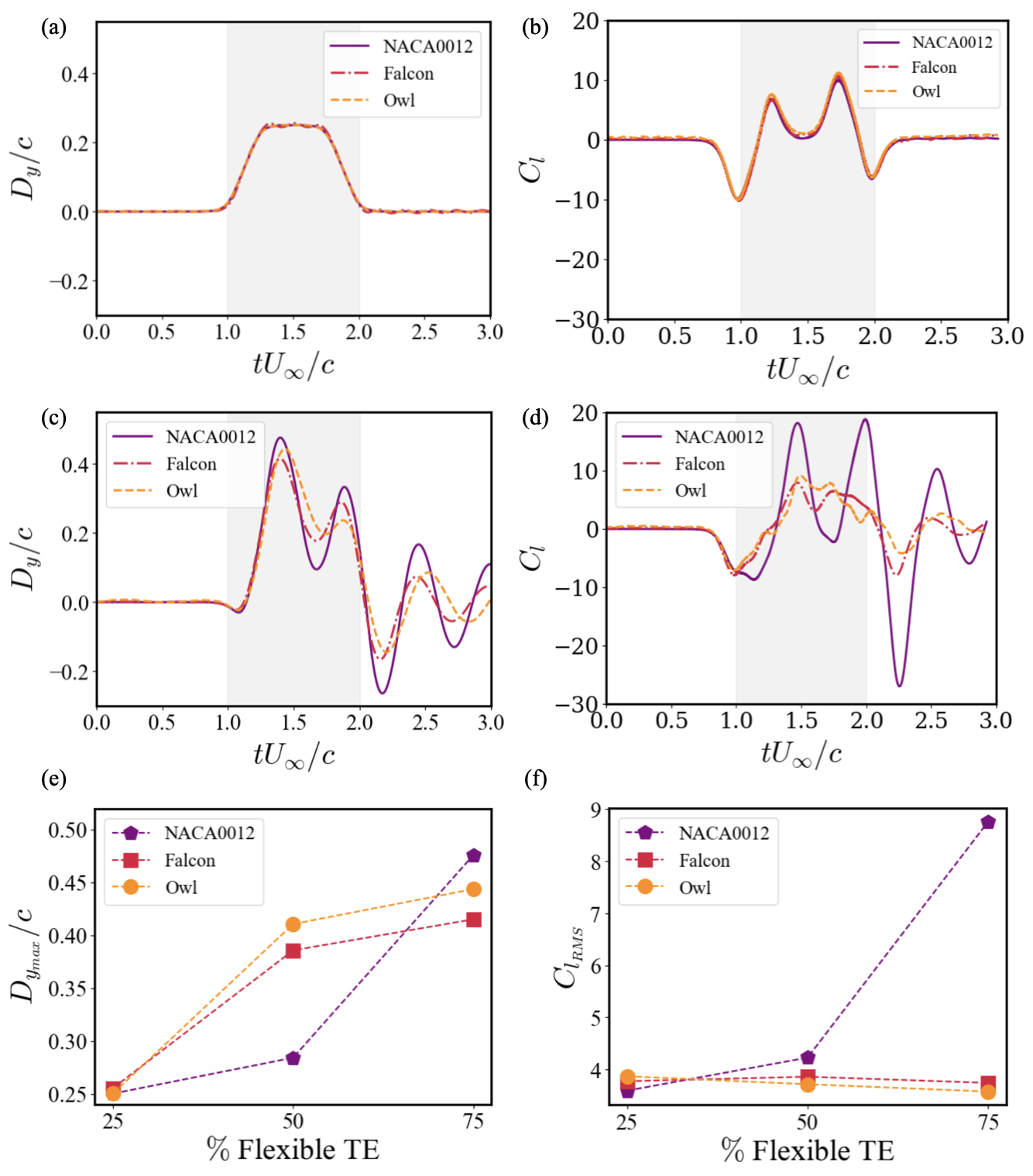}
    \caption{Time histories of $D_y/c$ and $C_l$ for the different wing geometries at $K_B = 5$ and $a_s = 11$ -- the $25\%$ flexible trailing-edge configuration (a, b) and the $75\%$ flexible trailing-edge configuration (c, d); the corresponding results for the $50\%$ case are presented in figure~\ref{figure: plunging - flexibility}(e, f). Sub-figures (e) and (f) present the variation of $D_{y_{\max}}/c$ and $C_{l_{RMS}}$, respectively, as functions of the chordwise flexible segment extent for all three wing geometries. The shaded grey region denotes the duration of the accelerated plunging manoeuvre.}
    \label{figure: plunging - changing flexible segment size}
\end{figure}

In this section, the effect of flexible segment size is investigated by comparing the response and lift characteristics of the three different wing sections with varying flexible trailing-edge segment sizes. In addition to the previously considered 50\% flexible configuration, cases with 25\% and 75\% flexible trailing-edge are examined for $K_B=5$ and $a_s=11$. Figure~\ref{figure: plunging - changing flexible segment size} shows the temporal evolution of $C_l$ and $D_y/c$ for the three wing geometries for 25\% and 75\% flexible trailing-edge configurations. 

\begin{figure}
    \centering
    \includegraphics[width=1\textwidth]{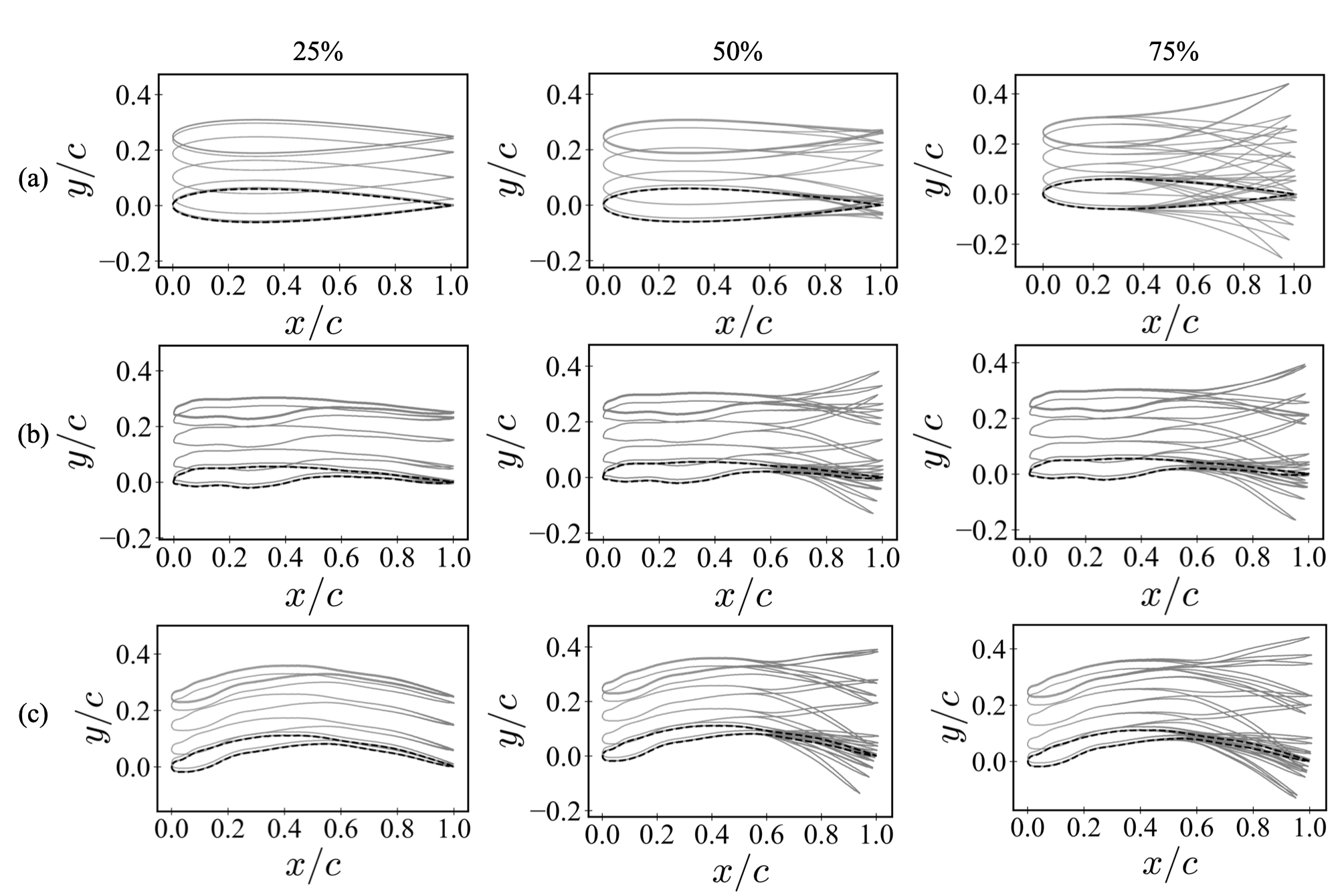}
    \caption{Displacement envelopes over the full gust manoeuvre for the (a) NACA0012 foil, (b) falcon, and (c) owl wing sections at $K_B = 5$ and $a_s = 11$, comparing three chordwise flexible segment extents: $25\%$ (left column), $50\%$ (centre column), and $75\%$ (right column) of the chord. Each grey line represents the instantaneous deformed wing profile at a discrete time instant, and the dashed black line denotes the undeformed reference geometry at the start. The progressive increase in both the spatial extent and amplitude of deformation with increasing flexible segment proportion is evident across all three wing geometries, with the owl wing section exhibiting the largest trailing-edge excursions at all flexibility levels.}
    \label{figure: displacement envelopes - segment size}
\end{figure}

\begin{figure}
    \centering
    \includegraphics[width=1\textwidth]{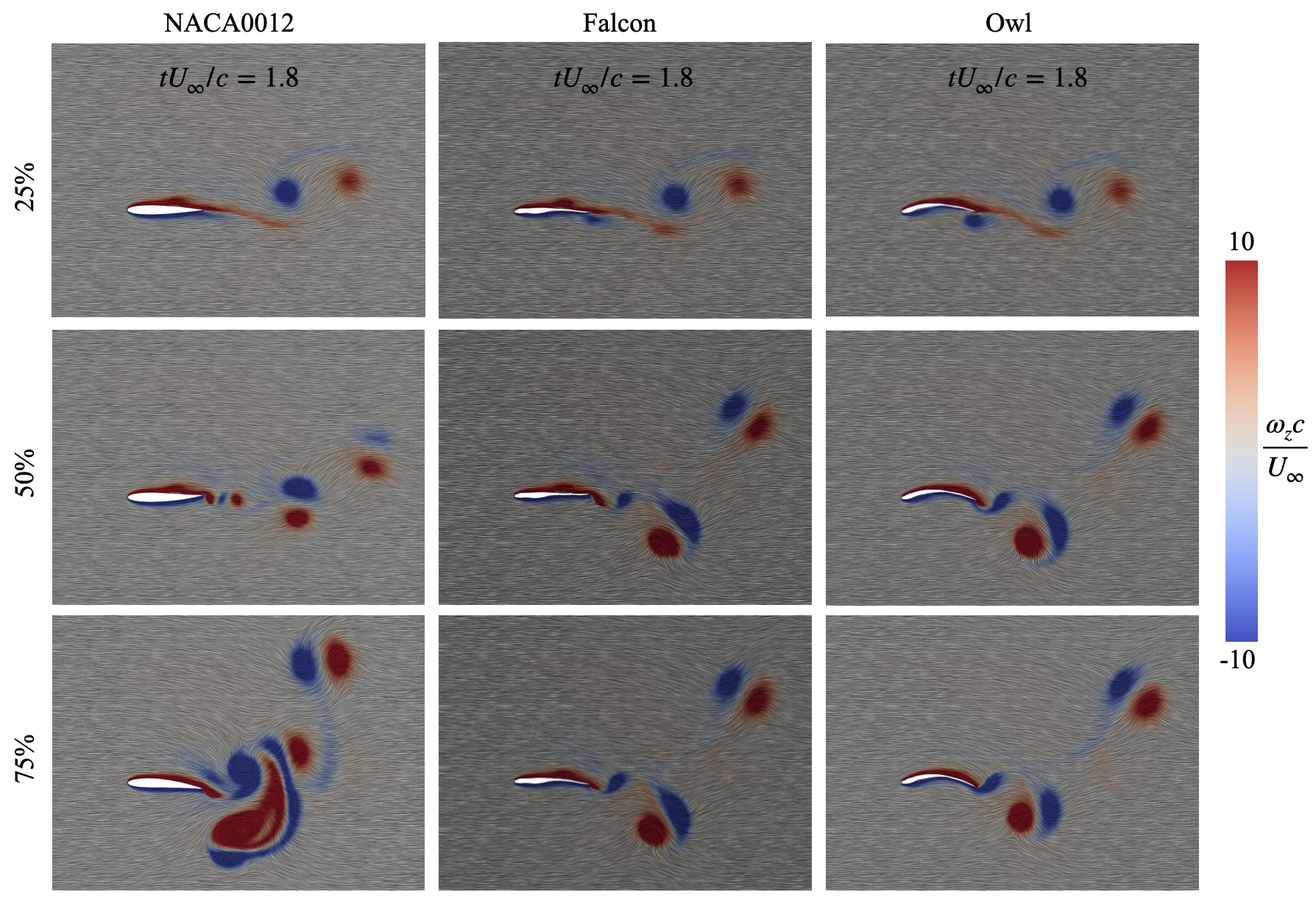}
    \caption{Instantaneous flow-field at $tU_\infty/c = 1.8$ for the NACA0012 foil (left column), falcon (centre column), and owl (right column) wing sections, comparing three chordwise flexible segment extents: $25\%$ (top row), $50\%$ (middle row), and $75\%$ (bottom row), all at $K_B = 5$ and $a_s = 11$. The flow topology is visualised using line integral convolution (LIC) overlaid with normalised spanwise vorticity contours ($-10 \le \omega_z c / U_\infty \le +10$) to highlight the near-wake vortical structures.}
    \label{figure: vorticity LIC - segment size}
\end{figure}

For all three wing sections, the $25\%$ trailing-edge flexible case exhibits minimal tip displacement (figure~\ref{figure: plunging - changing flexible segment size}(a)), with results closely resembling those of the $K_B = 100$ case presented in figures~\ref{figure: plunging - flexibility}(a, b). Similarly, the $C_l$ values are comparable across the cases, although slight deviations are observed in the peak values, with the bio-inspired wing sections exhibiting higher maximum values than the NACA0012 foil; see figure~\ref{figure: plunging - changing flexible segment size}(b). Increasing the length of the trailing-edge flexible segment leads to progressively greater trailing-edge tip displacement, with the $75\%$ flexible case exhibiting the most pronounced displacement; see figure~\ref{figure: plunging - changing flexible segment size}(c). For the 75\% flexible configuration, the lift response exhibits strong transient excursions associated with accelerated plunging, characterised by sharp peaks and troughs indicative of vortex-dominated loading; see figure~\ref{figure: plunging - changing flexible segment size}(d). Among all configurations, the $75\%$ flexible NACA0012 foil exhibits the largest trailing-edge deformation and displays the largest lift variability, with pronounced overshoots and deep negative excursions. In contrast, the bio-inspired wing sections (falcon and owl) exhibit visibly attenuated fluctuations, with smoother force evolution. 

Figure~\ref{figure: plunging - changing flexible segment size}(e) shows that the maximum deformation increases monotonically with the flexible trailing-edge segment size. However, the rate of increase differs across geometries. The NACA0012 foil exhibits a sharp increase in peak deformation at the 75\% flexible configuration, whereas the bio-inspired wing sections exhibit a more gradual increase. This indicates that curvature and thickness distributions redistribute structural compliance along the chord, moderating deformation sensitivity. The owl wing section maintains relatively high deformation even at moderate flexibility, suggesting more effective load redistribution. Figure~\ref{figure: plunging - changing flexible segment size}(f) reveals a marked contrast in lift variability. The NACA0012 foil exhibits a pronounced increase in $C_{l_{RMS}}$ with trailing-edge flexibility, indicating that increased compliance amplifies unsteady loading. Conversely, both bio-inspired wing sections exhibit nearly invariant RMS levels across variations in flexibility. This behaviour demonstrates that bio-inspired geometries suppress the growth of force fluctuations despite enhanced structural deformation. The results imply that the camber-induced modification of the leading-edge pressure gradient stabilises vortex evolution, mitigating irregular vortex-induced loading that inflates RMS values.

The displacement envelopes in figure~\ref{figure: displacement envelopes - segment size} illustrate the spatial character of chordwise flexibility. Across all three wing sections, the $25\%$ flexible cases exhibit negligible deformation, resembling rigid behaviour (as observed for $K_B = 100$), while the $50\%$ and $75\%$ flexible cases display progressively larger deflections. For $75\%$ flexible TE, the NACA0012 foil (see figure~\ref{figure: displacement envelopes - segment size}(a)) exhibits largely harmonic trailing-edge bending with limited curvature redistribution; refer to Supplementary video 1 for the visualisation of the displacement envelopes for 50\% and 75\% flexible trailing-edge configurations. The falcon wing section (see figure~\ref{figure: displacement envelopes - segment size}(b)) shows enhanced trailing-edge deflection with increased shape asymmetry, demonstrating stronger elastic participation. The owl wing section (see figure~\ref{figure: displacement envelopes - segment size}(c)) displays the most pronounced distributed curvature, suggesting that the structural deformation dynamically alters effective camber. These patterns confirm that wing geometry governs not only the magnitude of deformation but also its topology, which directly influences instantaneous aerodynamic behaviour. 

Figure~\ref{figure: vorticity LIC - segment size} presents snapshots of the flow field at $tU_\infty/c=1.8$ across the configurations (refer to Supplementary video 3 for the flow-field visualisation for 50\% and 75\% flexible trailing-edge configurations). The flow fields, visualised using line integral convolution (LIC) overlaid with vorticity contours, reveal changes in wake topology with increased trailing-edge flexibility. For the NACA0012 foil, greater flexibility ($75\%$) leads to large-scale vortex distortion and loss of coherence, consistent with the earlier-observed elevated lift fluctuations. The wake exhibits irregular vortex interactions indicative of increased flow instability. In contrast, the falcon and owl wing sections maintain more organised vortex structures even at high flexibility. The vortex cores remain coherent and exhibit smoother convection trajectories. This suggests that bio-inspired curvature/thickness distributions regularise the shear-layer roll-up and vortex-pairing processes. The owl wing section, in particular, shows a stable vortex couple with reduced wake distortion, consistent with its lower force variability.

Collectively, the figures demonstrate that increasing trailing-edge flexibility enhances structural deformation but does not uniformly amplify aerodynamic variability. The NACA0012 foil exhibits a direct coupling between flexibility and force fluctuations, whereas bio-inspired wing sections decouple deformation amplitude from force intermittency. This behaviour indicates a stabilisation mechanism wherein curvature and thickness distributions regulate vortex evolution, suppressing irregular vortex dynamics that drive lift variability. The present results suggest that passive morphing acts not merely as a compliant structural response but also as an effective geometric flow-control mechanism under transient accelerated conditions.

\section{Effect of transition speed parameter}
\label{sec:acceleration}

\begin{figure}
    \centering
    \includegraphics[width=1\textwidth]{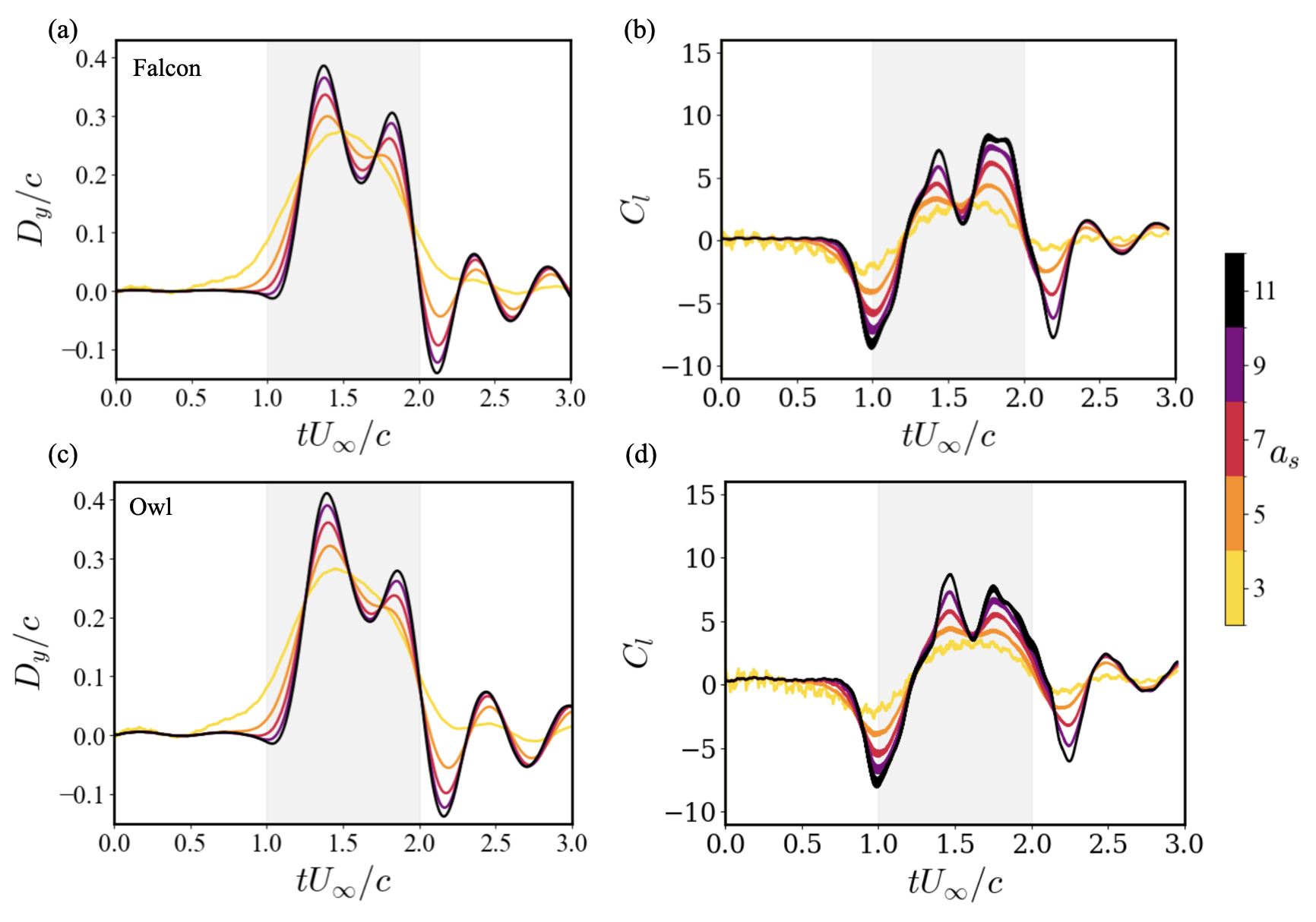}
    \caption{Time histories of $D_y/c$ and $C_l$ for the falcon (a, b) and owl (c, d) wing sections at $K_B = 5$, comparing the effect of the transition speed parameter at $a_s = 3$, $5$, $7$, $9$, and $11$, indicated by the colour scale from yellow (lowest) to black (highest). The shaded grey region denotes the duration of the accelerated plunging manoeuvre.}
    \label{figure: plunging - changing as}
\end{figure}

\begin{figure}
    \centering
    \includegraphics[width=1\textwidth]{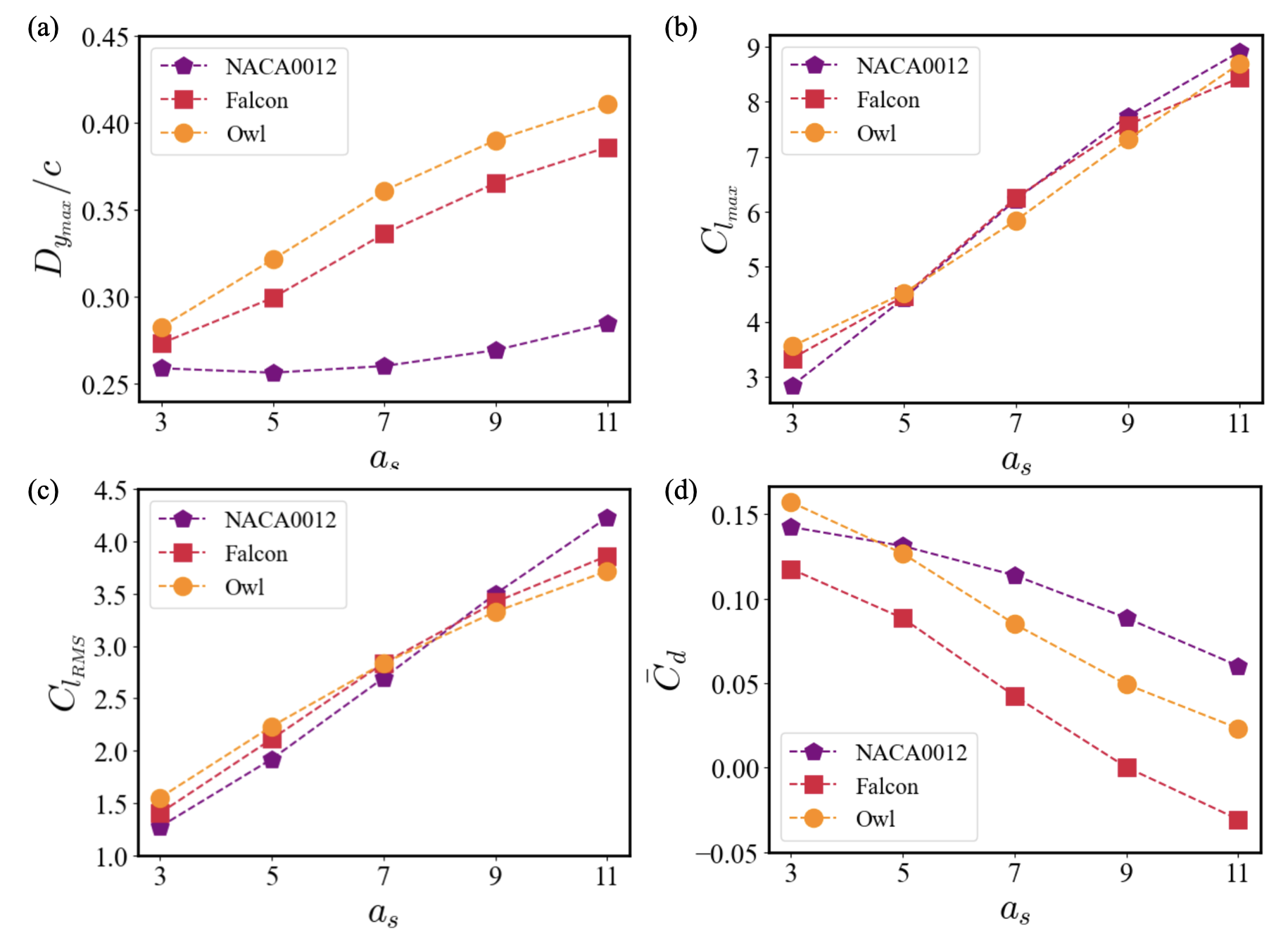}
    \caption{Comparison of structural deformation and aerodynamic metrics with varying $a_s$ for the NACA0012 foil, falcon, and owl wing sections at $K_B = 5$: (a) $D_{y_{\max}}/c$, (b) $C_{l_{\max}}$, (c) $C_{l_{RMS}}$, and (d) $\bar{C}_d$. All metrics are evaluated over the full simulation time window $tU_\infty/c = 0$ to $3.0$.}
    \label{figure: max values and RMS - as}
\end{figure}

\begin{figure}
    \centering
    \includegraphics[width=1\textwidth]{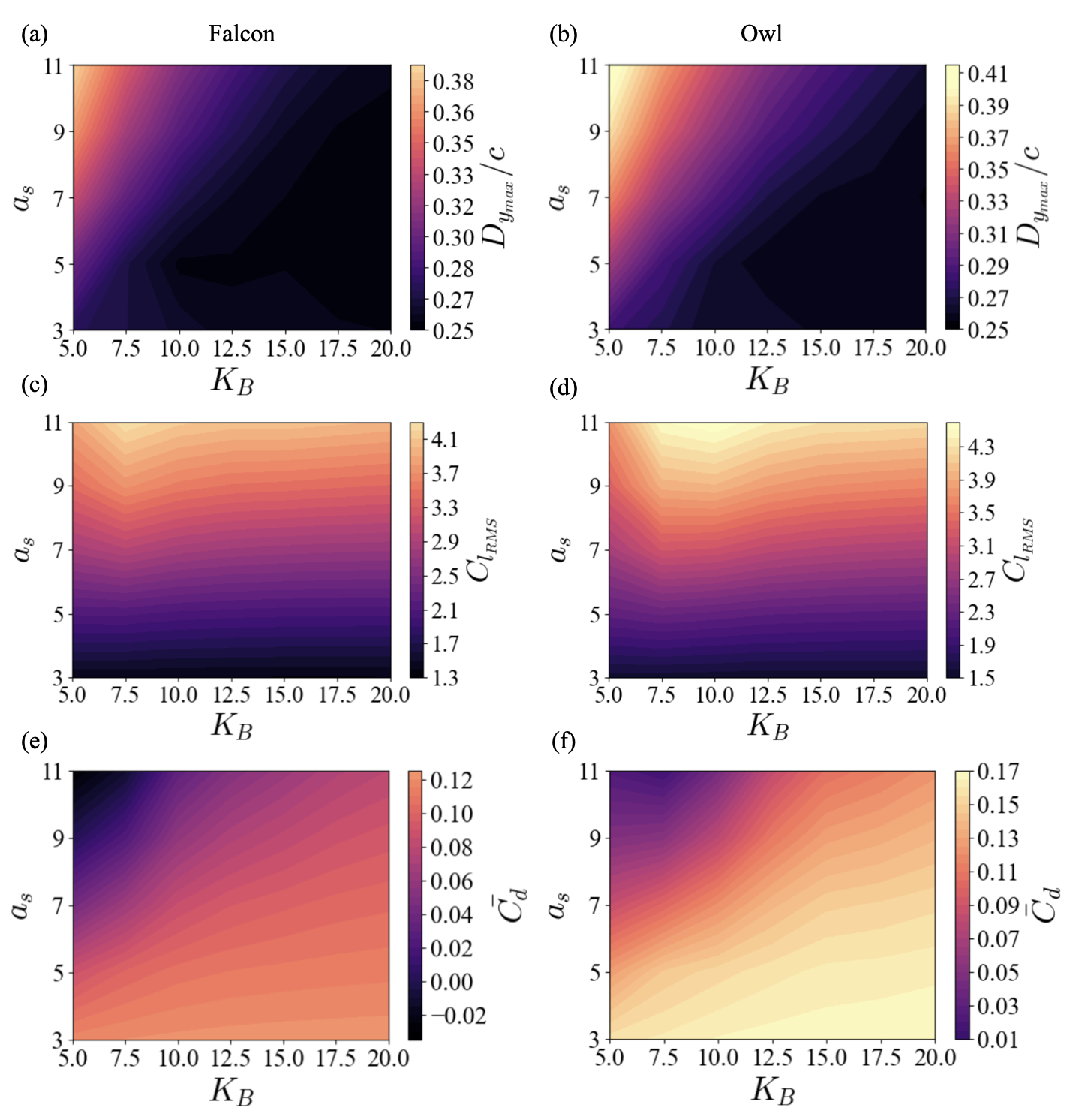}
    \caption{Contour maps of structural deformation and aerodynamic metrics in the $(K_B, a_s)$ parameter space for the falcon (left column) and owl (right column) wing sections: $D_{y_{\max}}/c$ (a, b), $C_{l_{RMS}}$ (c, d), and $\bar{C}_d$ (e, f). The contours are constructed from the discrete parameter combinations investigated.}
    \label{figure: contour maps}
\end{figure}

\begin{figure}
    \centering
    \includegraphics[width=1\textwidth]{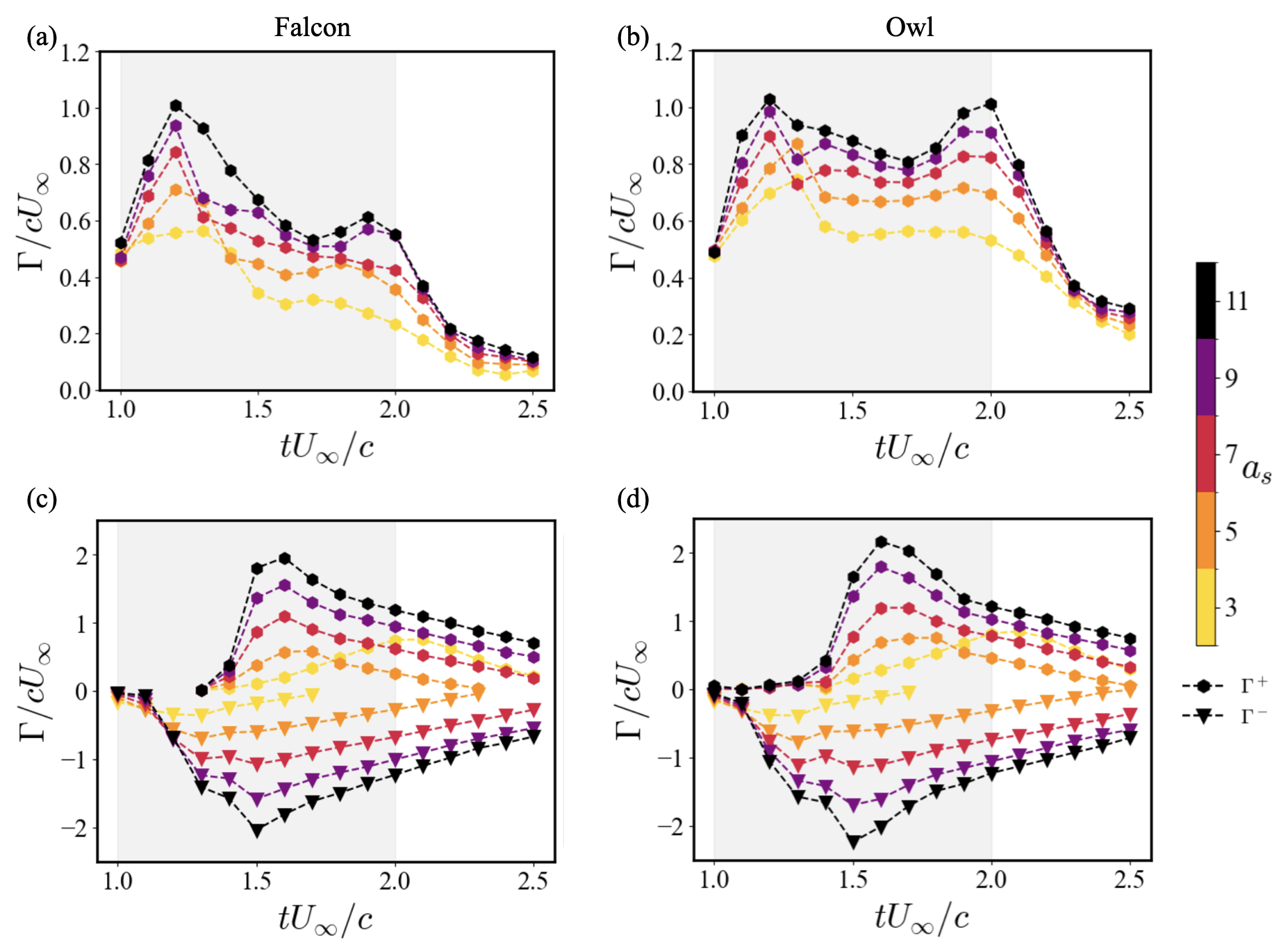}
    \caption{Comparison of the normalised circulation ($\Gamma/cU_\infty$) values with varying $a_s$ for LEV and TEVs over dimensionless time $tU_\infty/c=1$ to $2.5$ for falcon (a, c) and owl (b, d) wing sections at $K_B=5$. Positive circulation ($\Gamma^+$) and negative circulation ($\Gamma^-$) are marked by circle and inverted triangle markers, respectively. The line colours for different $a_s$ are indicated by the colour bar scale from yellow (lowest) to black (highest). The shaded grey region denotes the duration of the accelerated plunging manoeuvre.}
    \label{figure: circulation  - changing as}
\end{figure}

\begin{figure}
    \centering
    \includegraphics[width=1\textwidth]{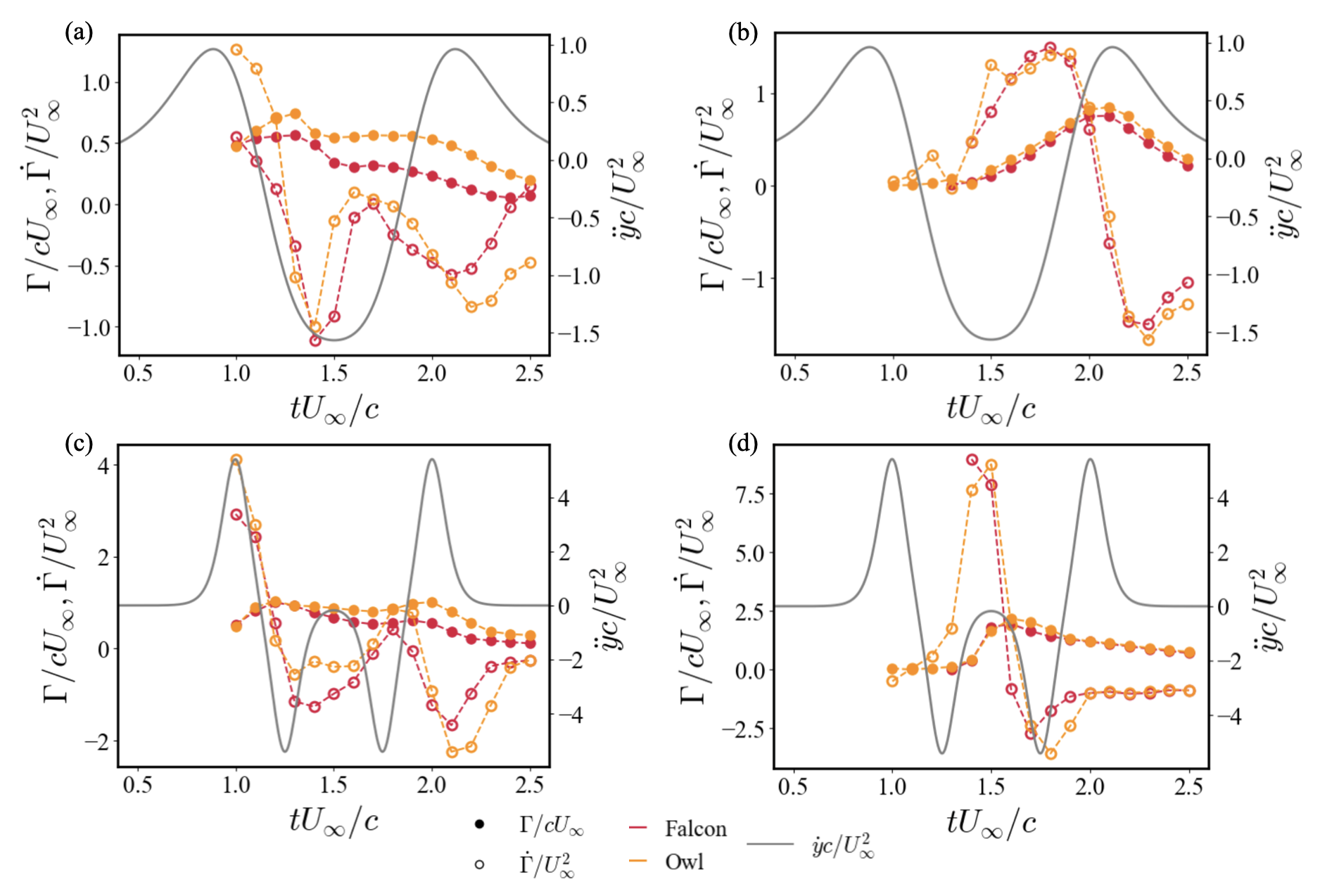}
    \caption{Comparison of $\Gamma/cU_{\infty}$ (filled circles) and $\dot{\Gamma}/U_{\infty}^2$ (open circles) for falcon (red) and owl (orange) wing sections, at $K_B = 5$. The left panels (a, c) show LEV and the right panels (b, d) show TEV quantities. The rows correspond to $a_s = 3$ (a, b) and $a_s = 11$ (c, d). The normalised plunge acceleration $\ddot{y}c/U_\infty^2$ (grey line, right axis) is included as a reference.}
    \label{figure: circulation comparison - changing as}
\end{figure}

In this section, the effect of mimicked gust strength on the FSI response is examined by varying the transition speed parameter $a_s$; five different values ($a_s=3, 5, 7, 9$, and $11$) are investigated. The case with $a_s=3$ represents a gradual kinematic transition, whereas $a_s=11$ represents the most impulsive acceleration considered. Figures~\ref{figure: plunging - changing as}(a, c) show that $D_y/c$ increases systematically with increasing $a_s$ for both the falcon and owl wing sections. Higher acceleration produces larger deformation amplitudes and more pronounced asymmetry, implying that transient inertial forcing enhances fluid-induced elastic response. The deformation peak shifts slightly in time as $a_s$ increases, reflecting acceleration-dependent elastic phase lag. Notably, larger accelerations amplify both the positive deformation excursion and the subsequent recoil, suggesting stronger coupling between the fluid forcing and structural dynamics. Figures~\ref{figure: plunging - changing as}(b, d) show the corresponding lift response. The lift peaks increase markedly with increasing $a_s$, consistent with the circulation trends (figure~\ref{figure: circulation  - changing as}). Higher accelerations produce stronger lift amplification, whereas lower accelerations yield comparatively weaker force amplification, despite producing non-negligible structural deformation. This decoupling between deformation amplitude and lift magnitude highlights the dominant role of vortex coherence.

Figure~\ref{figure: max values and RMS - as}(a) shows that $D_{y_{max}}/c$ increases monotonically with increasing $a_s$ for all configurations. The owl wing section exhibits the largest deformation across the entire range, followed by the falcon wing section, while the NACA0012 foil remains the least deformable. The approximately linear trend indicates that acceleration directly amplifies fluid-induced elastic response. The persistent separation between the curves confirms that geometric compliance distribution governs the deformation sensitivity in addition to kinematic acceleration. Figure~\ref{figure: max values and RMS - as}(b) shows that the maximum $C_l$ also increases monotonically with increasing $a_s$ for all configurations. However, $C_{l_{max}}$ values remain comparable for the three wing sections. Figure~\ref{figure: max values and RMS - as}(c) presents the RMS lift fluctuations. Increasing $a_s$ produces a systematic growth in $C_{l_{RMS}}$, reflecting intensified unsteady vortex-induced loading under stronger acceleration. The NACA0012 foil exhibits the steepest increase, whereas the bio-inspired wing sections exhibit comparatively moderate growth. This behaviour indicates that geometry-dependent deformation mediates force variability, partially suppressing acceleration-induced fluctuations. Figure~\ref{figure: max values and RMS - as}(d) shows the mean drag coefficient $\bar{C_d}$. Increasing $a_s$ leads to a systematic drag reduction for the falcon and owl wing sections, while the NACA0012 foil exhibits a relatively lower decrease rate. At higher acceleration, the falcon wing section transitions to slightly negative drag, demonstrating net thrust generation. This behaviour reflects acceleration-driven redistribution of wake momentum, enhanced by bio-inspired geometry.

Flexible wing sections can respond to aerodynamic loading by increasing deformation, thereby exploiting their passive gust-mitigation mechanisms. Figures~\ref{figure: contour maps}(a, b) illustrate the deformation landscape in the $(K_B, a_s)$ parameter space. In all cases, deformation decreases with increasing stiffness (with higher $K_B$), consistent with structural scaling. In contrast, deformation increases strongly with $a_s$. The contours highlight a pronounced sensitivity at low $K_B$, where small stiffness variations produce large deformation changes. The owl wing section maintains elevated deformation levels over a wider stiffness range, indicating enhanced deformation robustness. For $K_B=20$, the displacements remain low and exhibit minimal variation across all values of $a_s$. This behaviour suggests that, for relatively stiffer wing sections, structural resistance suppresses deformation regardless of the $a_s$ value. In contrast, at low $K_B$ for the more flexible cases, the contours exhibit gradients with respect to $a_s$, indicating increased sensitivity to $a_s$. The largest deformation occurs for the most flexible configuration ($K_B=5$) at $a_s=11$. For this flexible case, $D_{y_{max}}$ increases monotonically with $a_s$. The contours reveal a non-linear interaction between $K_B$ and $a_s$, with the influence shifting from transition speed-dominated behaviour at low $K_B$ to stiffness-dominated behaviour at high $a_s$. Although the falcon and owl wing sections exhibit qualitatively similar displacement patterns, the owl wing section reaches higher values of $D_{y_{max}}$ in the parametric space. This suggests that both wing sections react with the same underlying mechanism, but geometric differences amplify the owl's structural response. 

Figures~\ref{figure: contour maps}(c, d) show the lift fluctuation behaviour in the $(K_B, a_s)$ space. Lift fluctuations increase primarily with $a_s$, while stiffness exerts a secondary stabilising influence. Higher $K_B$ reduces fluctuation intensity, indicating that structural rigidity attenuates vortex-induced variability. For all stiffness values $C_{l_{RMS}}$ increases with $a_s$. However, both wing sections exhibit distinct maxima at intermediate $K_B=7.5$ and $10$ for the falcon and owl, respectively, and at $a_s=11$. This implies that the configurations that maximise lift do not coincide with those that maximise displacement. At very low $K_B$, although deformation is substantial, excessive flexibility likely compromises aerodynamic efficiency and limits lift enhancement. Conversely, at high $K_B$, deformation is insufficient to fully exploit the flexibility. The optimal region, therefore, represents a balance between structural compliance and aerodynamic performance. Figures~\ref{figure: contour maps}(e, f) further show the drag landscape. $\bar{C_d}$ decreases with increasing $a_s$ and increases with stiffness, revealing a competition between transient inertial forcing and structural rigidity. Low-stiffness regimes promote drag reduction and thrust generation, particularly for the falcon wing, whereas high stiffness suppresses these effects. Overall, figure~\ref{figure: contour maps} demonstrates that acceleration is a key driver of deformation and unsteady loading, while stiffness governs the response amplitude. Bio-inspired geometries modify both deformation sensitivity and aerodynamic variability, thereby enhancing deformation capability, moderating lift fluctuations, and improving drag performance. These trends confirm that vortex-mediated force generation under accelerated motion is governed by a coupled acceleration–stiffness–geometry interaction.


Figure~\ref{figure: circulation  - changing as} presents the temporal variation of the total circulation magnitude for both bio-inspired wing sections, for the LEV (see figures~\ref{figure: circulation  - changing as}(a, b))  and the TEVs (see figures~\ref{figure: circulation  - changing as}(c, d)). Increasing $a_s$ results in a clear increase in circulation throughout the cycle, consistent with the larger structural deformation observed previously in figure~\ref{figure: plunging - changing as}. 
This behaviour indicates that increased acceleration enhances vorticity production and promotes the roll-up of stronger vortex structures. Both the positive and negative circulation components, $\Gamma^{+}$ and $\Gamma^{-}$, increase in magnitude with increasing $a_s$, with the change more pronounced for $\Gamma^{+}$. These results suggest that strong acceleration promotes the growth of dominant coherent vortices, leading to a more symmetric and stronger vorticity distribution. As $a_s$ increases, deformation is enhanced through stronger transient forcing, while vortex coherence is simultaneously strengthened, resulting in increased circulation and lift production (figure~\ref{figure: plunging - changing as}). 

\begin{figure}
	\centering
	\includegraphics[width=1\columnwidth]{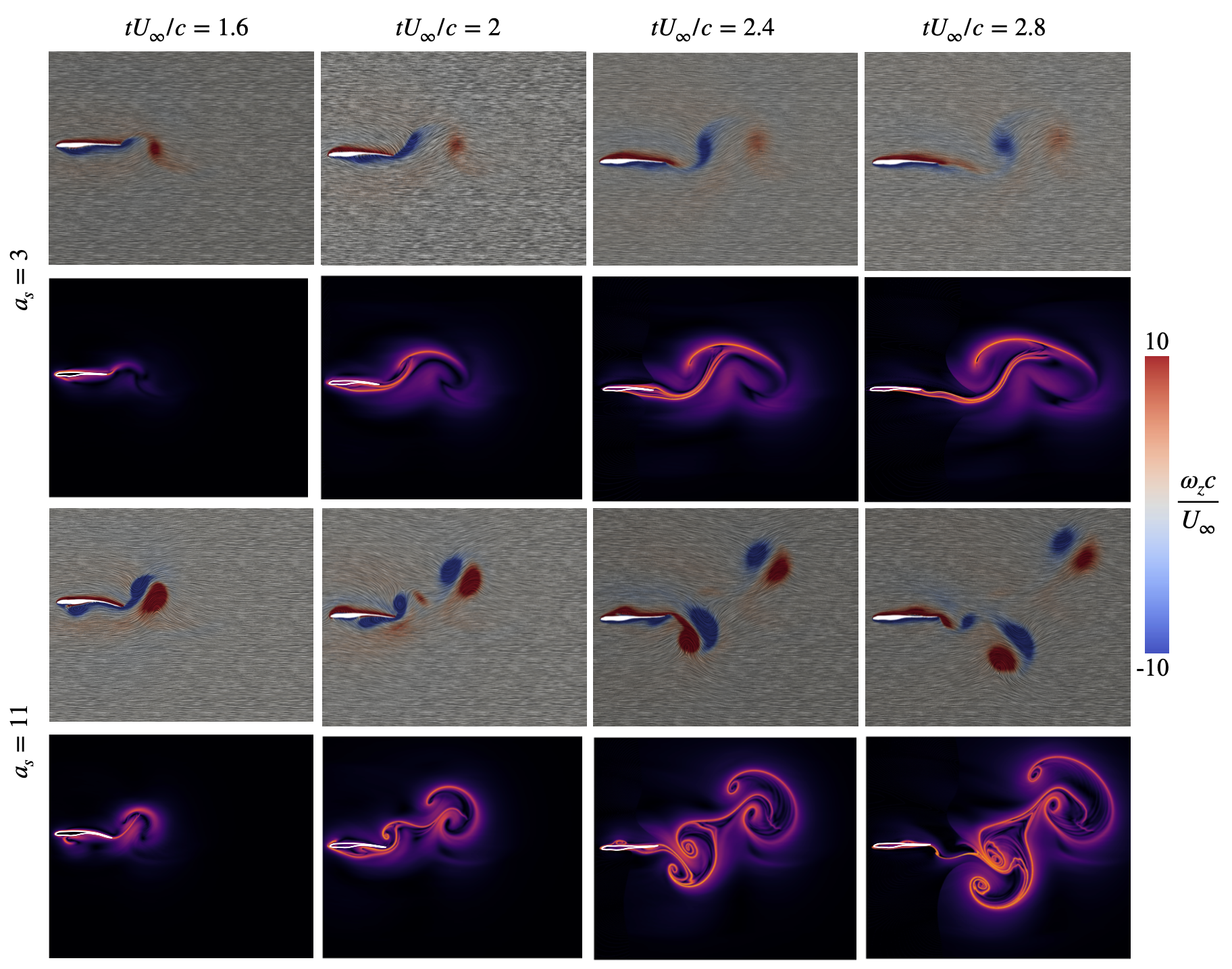}
	    \caption{Vorticity fields and Lagrangian flow structures for the falcon wing section at $K_B = 5$, comparing $a_s = 3$ (rows 1--2) and $a_s = 11$ (rows 3--4) at four representative time instances $tU_{\infty}/c = 1.6,\ 2.0,\ 2.4$, and $2.8$. Odd rows show normalised vorticity ($-10 \le \omega_z c/U_{\infty} \le +10$) contours (red: positive/counter-clockwise; blue: negative/clockwise) overlaid on line integral convolution (LIC) streamlines, which visualise instantaneous flow topology. Even rows show the corresponding backward finite-time Lyapunov exponent (FTLE) fields, representative of material barriers in the wake structures, where high values (bright regions) identify Lagrangian coherent structures (LCS).}
    \label{figure: fig6}
\end{figure}

Figure~\ref{figure: circulation comparison - changing as} presents the sequential vortical evolution for the bio-inspired wing sections at the lowest and highest transition speed parameters investigated, $a_s = 3$ (a, b) and $a_s = 11$ (c, d). At the lower acceleration ($a_s = 3$), the LEV forms and subsequently convects downstream while gradually weakening. This is followed by the development of a TEV, which also weakens as it convects, with the negative TEV dissipating particularly rapidly. In contrast, the higher acceleration case ($a_s = 11$) exhibits markedly different vortex dynamics. Rapid kinematic variation promotes stronger vorticity generation and sustained circulation, which subsequently decays gradually as the manoeuvre progresses. The TEVs remain significantly stronger throughout the plunging manoeuvre. The comparison demonstrates that increasing $a_s$ fundamentally amplifies vortex strength and produces more coherent wake structures, as also shown in figure~\ref{figure: circulation  - changing as}. Higher acceleration is associated with more coherent LEV–TEV coupling and more organised wake development, whereas the lower acceleration case exhibits earlier dissipation. These observations are consistent with the circulation and force trends, confirming that acceleration influences vortex coherence and stability, which in turn affect aerodynamic force production.

Figure \ref{figure: fig6} compares the unsteady flow field around the falcon wing section at the lowest ($a_s=3$) and highest ($a_s=11$) transition-speed parameters at $tU_\infty/c=1.6$. As presented before, LIC visualisation, overlaid with vorticity contours, provides a simultaneous view of the instantaneous streamline topology and the evolution of vortical structures. Additionally, the backward finite-time Lyapunov exponent (FTLE) ridges reveal Lagrangian coherent structures, which are representative of material barriers in the unsteady flow field (see Supplementary video 4). At $tU_\infty/c=1.6$, both the positive LEV and the TEVs are significantly stronger for $a_s=11$ as compared to that observed for $a_s=3$. For $a_s=3$, the LEV is less distinct and remains largely attached to the wing surface, whereas the LEV is seen to be considerably stronger for the $a_s=11$ cases. This observation is also consistent with the circulation values for the LEV presented in figure~\ref{figure: circulation  - changing as}. Furthermore, the $a_s=3$ case exhibits relatively weak TEV formation. In contrast, the $a_s=11$ case shows the formation and subsequent shedding of strong, coherent trailing-edge vortex couples in the wake. 

\section{Concluding remarks}
\label{sec:conclusions}

The present study investigates the FSI response of avian-inspired passively morphing foils subjected to accelerated plunging, demonstrating that plunging acceleration, chordwise flexibility distribution, and wing geometry govern both structural deformation and vortex-mediated force generation in a coupled manner, in contrast to rigid wing sections. Among the flexible trailing-edge configurations investigated, the optimal bending stiffness is found to be geometry-dependent: the $K_B = 10$ case yielded the highest $C_l$ for both the NACA0012 foil and owl wing sections, whilst the falcon wing section achieved peak performance at $K_B = 7.5$, with all flexible configurations outperforming the stiffer ($K_B = 100$) counterparts. This geometry dependence of the optimal stiffness underscores the need to tailor material properties to specific wing geometries rather than applying a universal compliance criterion.

Of the three geometries examined, the owl wing section consistently produces the highest transverse displacement $D_y/c$ and lift coefficient $C_l$, together with the strongest and most coherent leading- and trailing-edge vortical structures, followed by the falcon wing section and the NACA0012 foil. When the chordwise extent of the flexible segment is varied, the $75\%$ flexible configuration exhibits very high deformation, resulting in a highly fluctuating lift response, whilst the $25\%$ case behaves near-identically to a rigid wing. Notably, the $C_{l_{RMS}}$ for the $75\%$ flexible NACA0012 foil shows a rapid jump compared to the $50\%$ flexible case. On the contrary, the $C_{l_{RMS}}$ reduces slightly for the bio-inspired wing sections as the chordwise extent of the flexible segment increases from $25\%$ to $75\%$.

Varying $a_s$ revealed a monotonic relationship between acceleration magnitude and both structural deformation and vortex strength. The highest acceleration ($a_s = 11$) induced the largest trailing-edge deflections and the most pronounced leading- and trailing-edge vortices, whilst the lowest ($a_s = 3$) produced minimal deformation and weaker vortex interactions, closely resembling the response of a rigid wing section. Across all cases, the vortices generated during the gust interaction played a central role in driving trailing-edge deflection and modifying the effective camber of the wing, directly influencing its instantaneous aerodynamic performance.

Note that the present study is based on two-dimensional simulations, conducted at a fixed density ratio of $\rho^* = 100$ and in the low Reynolds number regime ($Re = 10^3$). Extending the parameter space to a wider range of material properties would enable the identification of geometry-specific optimal configurations, whilst three-dimensional simulations would capture spanwise-flow phenomena and tip effects. Future investigations at Reynolds numbers representative of birds and bio-inspired aerial vehicles ($Re = \mathcal{O}(10^4)$--$\mathcal{O}(10^5)$) are necessary to determine how turbulent boundary-layer development and Reynolds-number-dependent separation affect the optimal stiffness distributions reported. 

More fundamentally, the assumption of discrete rigid-to-flexible transitions along the chord represents a considerable simplification of avian wing morphology. Incorporating spatially graded compliance, enabling smooth stiffness transitions from a rigid leading-edge to a compliant trailing region, would more accurately replicate the structural mechanics of biological lifting surfaces and provide a stronger physical basis for the aerodynamic benefits observed here. 

\section*{Acknowledgment}

The authors gratefully acknowledge the support of the UK Engineering and Physical Sciences Research Council (EPSRC) for the DLA scholarship of Hibah Saddal, grant EP/W524396/1. This research is also partially supported by the Royal Society International Exchanges Funding, Grant No. IES$\backslash R2 \backslash 242343$ (PI: Chandan Bose). The present FSI simulations are carried out using the computational resources provided by the UK national supercomputing facility ARCHER2 through the EPSRC Access To HPC Pioneer Grant and the UK Turbulence Consortium (PI: Chandan Bose). 

\section*{Supplementary materials}

Supplementary video 1 illustrates the time evolution of structural displacement envelopes over the course of the gust manoeuvre for the three wing geometries -- NACA0012, falcon, and owl -- at $K_B = 5$, comparing the lowest and highest transition speeds considered ($a_s = 3$ and $a_s = 11$), as well as for the $75\%$ flexible trailing-edge configuration. Supplementary video 2 shows the vorticity contour fields used in the circulation analysis, with the tracked leading- and trailing-edge vortices explicitly identified, for $K_B = 5$ and $a_s = 11$ across all three wing geometries. Supplementary video 3 presents the corresponding unsteady flow fields, visualised using line integral convolution (LIC) and overlaid with vorticity contours. Supplementary video 4 presents backward finite-time Lyapunov exponent (FTLE) ridge fields for the falcon wing section at $K_B = 5$, and contrasts the topological differences in these structures between the $a_s = 3$ and $a_s = 11$ cases.

\section*{Deceleration of Interests}

The authors report no conflict of interest.

\bibliographystyle{jfm}
\bibliography{jfm}

\end{document}